\documentclass[aps,pra, twocolumn, groupedaddress, twoside]{revtex4-2}

\usepackage{graphicx, amsfonts, amsmath, amssymb}

\usepackage{xcolor}
\definecolor{grey}{rgb}{0.7,0.7,0.7}
\definecolor{db}{rgb}{0,0,0.5}

\def\bse{\begin{subequations}}
\def\ese{\end{subequations}}
\def\beqn{\begin{eqnarray}}
\def\eeqn{\end{eqnarray}}
\def\benum{\begin{enumerate}}
\def\eenum{\end{enumerate}}
\def\eq{equation}

\def\1i{\mathfrak{i}\hspace{1pt}}
\def\ft{\mathfrak{t}}
\def\asin{\sin^{-1}}

\def\atan{\tan^{-1}}
\def\xhat{\hat{\mathbf{x}}}
\def\yhat{\hat{\mathbf{y}}}
\def\zhat{\hat{\mathbf{z}}}

\def\dhat{\hat{\mathbf{d}}}

\def\hxhat{\hspace{1.5pt}\hat{\mathbf{x}}}
\def\hyhat{\hspace{1.5pt}\hat{\mathbf{y}}}
\def\hzhat{\hspace{1.5pt}\hat{\mathbf{z}}}
\def\hehat{\hspace{1.5pt}\hat{\mathbf{e}}}
\def\sighat{\hat{\boldsymbol{\sigma}}}
\def\hsighat{\hspace{1.5pt}\hat{\boldsymbol{\sigma}}}
\def\rt2{\sqrt{2}}

\def\bk{\mathbf{k}}
\def\bE{\mathbf{E}}
\def\bcE{\boldsymbol{\mathcal{E}}}
\def\tbE{\tilde{\mathbf{E}}}

\def\cE{\mathcal{E}}

\def\cI{\mathcal{I}}
\def\cP{\mathcal{P}}

\def\bRhat{\hat{\mathbf{R}}}

\def\roc{radius of curvature }

\def\wrt{with respect to }

\def\thi0{\theta_{i0}}

\def\cth{\cos\theta}
\def\sth{\sin\theta}
\def\cph{\cos\phi}
\def\sph{\sin\phi}
\def\ccph{\cos^2\phi}
\def\ssph{\sin^2\phi}

\def\circf{\mathrm{circ}}

\def\mum{$\mu$m}
\def\mums{$\mu$m }

\def\deg{^\circ}
\def\hpt{\hspace{1pt}}
\def\hppt{\hspace{1.5pt}}
\def\hem{\hspace{1em}}
\def\hemm{\hspace{1.5em}}

\usepackage{fancyhdr}
 
\begin{document}



\title{\colorbox{db}{\parbox{\linewidth}{  \centering \parbox{0.9\linewidth}{ \textbf{\centering\Large{\color{white}{ \vskip0.2em{Complex Far-Fields and Optical Singularities due to Propagation Beyond Tight Focusing: Combined Effects of Wavefront Curvature and Aperture Diffraction}\vskip0.8em}}}}}}}




\author{Nitish Kumar}
\affiliation{School of Physics, University of Hyderabad, Hyderabad 500046, India}

\author{Anirban Debnath}
\email[]{anirban.debnath090@gmail.com}
\affiliation{School of Physics, University of Hyderabad, Hyderabad 500046, India}

\author{Nirmal K. Viswanathan}
\affiliation{School of Physics, University of Hyderabad, Hyderabad 500046, India}


\date{\today}

\begin{abstract}
\noindent{\color{grey}{\rule{0.784\textwidth}{1pt}}}
\vspace{-0.8em}

All optical systems, which involve the collimation of a reflected, transmitted or scattered wave subsequent to tight focusing, are subject to two kinds of deviations. One is the wavefront curvature due to inaccurate focal placement of the interface or scatterer particle under consideration, and the other is the diffraction caused by the finite lens aperture. 
In the present paper we explore these phenomena in detail by considering a rigorous simulated model and an appropriate experimental setup. We hence demonstrate the complicated intensity profiles and optical singularity characteristics of the observed far field. 
Then we describe ways to minimize these deviations in a general experiment. 
But more importantly, our analysis proves that these deviations by themselves are significant optical phenomena of fundamental interest. The observed complex field profiles have similarities to standard diffraction-limited tight focal fields, though our field detection is different from the standard schemes. This indicates the relevance of these complex fields to a larger class of systems involving wavefront curvature and aperture diffraction. The detailed analysis and results of the present paper already serve as core explorations of these optical phenomena; 
and we also suggest future research directions where these system aspects can be purposefully created and explored further.
{\color{grey}{\rule{0.784\textwidth}{1pt}}}
\end{abstract}


\maketitle



\tableofcontents

{\color{grey}{\noindent\rule{\linewidth}{1pt}}}



\section{Introduction} \label{Sec_Intro}

The tight focusing of an optical beam due to an aplanatic high numerical aperture (NA) lens, e.g. a microscope objective, is a widely studied research area in optics. The first rigorous theory on the vector nature of the electromagnetic field at a tight focus was given by Richards and Wolf \cite{Wolf1959, RichardsWolf1959}. Subsequently, this phenomenon and its applications have been extensively explored in various optical systems, comprehensive discussions on which can be found, e.g., in Refs. \cite{Stamnes} and \cite{NanoOpticsBook}.


There are subclasses of systems, where a normal interface (\wrt the central propagation direction) or a scatterer particle is placed at or near the focus \cite{NanoOpticsBook, WilsonConfocal1997, TorokConfocal1998, InamiConfocal2000, NovotnyFocusImage2001, HelsethMagOptic2002, PetrovFocusedBeamRT2005, BerciaudPhHeterodyne2006, LermeSpectro2008, ZuchnerAuScatt2008, HuDarkFieldMicro2008, HuangNanoparticle2008, BanzerNano2010, RodriguezNanoprobe2010, KnightDarkFieldMicro2010, Bliokh2011, ZuchnerDonutRev2011, FanDarkFieldMicro2012, Brody2013, Papaioannou2018, VortexBrewster, Eismann2022, UB_NK_NKV_OL_2022, ADetalNormInc2023}. Some of these systems intend to study the scattered or transmitted wave after collecting or collimating it via another high NA lens. Some other systems intend to study the normal-reflected beam after collimating it back through the initial high NA lens. 
In any case, since the length scale of the focal region is significantly smaller than the other experimental length parameters, it is very difficult to place the interface or particle at the focus (or at a specific distance away from the focus) with sufficient accuracy. Any deviation from the intended placements of the optical system components disturbs the eventual collimation of the output beam. One can still consider an approximate collimation if the deviation is of the order of a wavelength. 
This has been considered, for example, in the normal reflection system described by Novotny et al. \cite{NovotnyFocusImage2001, NanoOpticsBook}, in which no wavefront curvature effect has been taken into account.
But deviations larger than $\sim10$ \mums (e.g., the smallest division of a typical translation stage) disturb the system sufficiently such that the final output beam cannot be considered as collimated. To our knowledge, the wavefront curvature effects due to such large displacements have not been reported in the literature.

An additional deviation from an ideal scenario is caused by the finite size of the lens aperture. It is known that the short-distance light propagation from the objective to the eyepiece of a compound microscope is efficiently approximated via geometrical ray tracing \cite{Hecht}. 
However, a general optical system such as the ones mentioned above can involve propagation to large enough distances (as compared to the size of the lens aperture) where aperture diffraction effects in the final observed beam field become considerable.
These effects are especially significant for a microscope objective, because it is a combination of lenses, each having a radius of a few millimeters, which is comparable to a typical beam width. 
In fact, it is a common practice in the above-mentioned class of optical systems to overfill the lens aperture in order to utilize the full NA of the aplanatic lens \cite{NanoOpticsBook}. But the very act of overfilling the aperture causes the beam to be abruptly blocked beyond the effective aperture boundary, thus causing aperture diffraction. This is the core reason why many experimental works in the literature show beam field profiles with Airy ring characteristics. These effects are further strongly enhanced due to the wavefront curvature.
In the presence of these deviations, it becomes particularly difficult to decipher which output field characteristics are caused by the interface properties or scattering phenomena under consideration, and which ones are caused by the deviations. So it is necessary to explicitly find out which effects in the beam field happen solely due to these deviations.

We address the above problem in the present paper. We first establish a simulated optical system in Sec. \ref{Sec_System}, 
where we rigorously model the wavefront curvature and aperture diffraction.
Then in Sec. \ref{Sec_Sim} we describe simulated results for observed intensity profiles considering different cases for different input parameter values. Remarkable optical singularity characteristics are observed for high convergence and divergence cases, which we explore in extensive detail. 
We then introduce an appropriate experimental setup in Sec. \ref{Sec_Exp} considering a tightly focused and subsequently reflected beam. The demonstrated experimental results agree well with the simulated results, thus verifying the correctness of our analysis and understanding of the problem.

With the knowledge of these deviation phenomena, one either can take appropriate steps to minimize these effects in a differently purposed experiment, or can interpret the sought after results even from the deviated field. 
But at this point, the attention of the discussion shifts to a new perspective. The simulated and experimentally demonstrated field profiles and optical singularities, though are complicated in nature, are described by precise mathematical expressions and posses well-recognizable patterns. 
These patterns have remarkable similarities with standard diffraction-limited tightly-focused fields, even though the final observed field at the detector of our system is not necessarily tightly focused. This indicates towards the possibility of appearance of these field patterns in more general contexts,
and we thus recognize that the considered deviation processes are significant electromagnetic optical phenomena by themselves. 
Our optical system modelling [Sec. \ref{Sec_System}], simulated field profile characteristics [Sec. \ref{Sec_Sim}] and experimental observations [Sec. \ref{Sec_Exp}] already serve as in-depth explorations of these optical phenomena and complex field profiles at a fundamental level. 
But looking forward, we also list a few research directions [Sec. \ref{Sec_Future}] along which one can explore the spin-orbit coupling (SOC) characteristics, optical singularity dynamics and various possible applications of these significant optical processes in the future.

\section{The Modeled Optical System} \label{Sec_System}

\begin{figure}
\includegraphics[width = \linewidth]{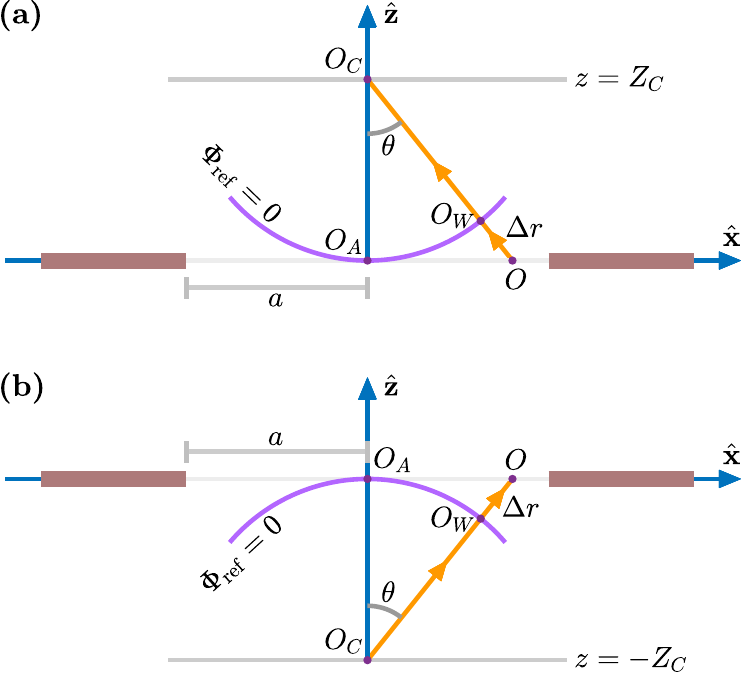}
\caption{
Simulated model configurations for (a) the converging case, (b) the diverging case. Here, $z = 0$ is the plane of the aperture. A wavefront with a radius of curvature $Z_C$ is shown in each case, where the $\Phi_\mathrm{ref} = 0$ reference phase is set.
}
{\color{grey}{\rule{\linewidth}{1pt}}}
\label{Fig_Model}
\end{figure}

The description of the problem [Sec. \ref{Sec_Intro}] indicates the requirement of an optical system model, where a curved wavefront beam, either converging or diverging, is passed through a circular aperture. We introduce such a model in Fig. \ref{Fig_Model}, considering the converging and diverging cases individually. In this model, the central ray of the beam propagates along the $z$ axis; and a circular aperture of radius $a$ is placed at the $z = 0$ plane coaxially to the beam.
In the converging case, the wavefronts appear to be converging towards a point $O_C$ at the $z = Z_C > 0$ plane [Fig. \ref{Fig_Model}(a)]; whereas, in the diverging case, they appear to be diverging from a point $O_C$ at the $z = -Z_C < 0$ plane [Fig. \ref{Fig_Model}(b)].
We use $Z_C$ as a known parameter in the model to control the wavefront curvature.

\subsection{Converging and Diverging Field Amplitudes} \label{Subsec_ConvDivEModel}

We now consider an initial condition that, in the absence of any wavefront curvature and aperture, the field amplitude vector at the $z = 0$ plane would be $\cE_0 \yhat$, where $\cE_0$ is a real field amplitude function. Such a simple condition is a necessity, to ensure that the final field characteristics we are to observe are solely due to the concerned deviations and not due to any complicated initial field characteristics.

Next we consider that, in the presence of wavefront curvature, the field amplitude at the $z = 0$ plane just before the aperture is a transformed form of $\cE_0 \yhat$, given by \cite{NanoOpticsBook, ADNKVrt2020}
\bse \label{R_operator}
\beqn
& \bcE' = g \hppt \bRhat^{-1}(\phi) \hpt \bRhat(\alpha\theta) \hpt \bRhat(\phi) \hppt \cE_0 \hyhat; & \\
& \mbox{where,} \hem \theta = \atan\left(\dfrac{\rho}{Z_C}\right) \! , \hem g = \dfrac{1}{\sqrt{\cth}}, & \\ 
& \bRhat(\phi) = \begin{bmatrix}
\cph & \sph & 0 \\ -\sph & \cph & 0 \\ 0 & 0 & 1
\end{bmatrix} \!, & \\
& \bRhat(\theta) = \begin{bmatrix}
\cth & 0 & -\sth \\ 0 & 1 & 0 \\ \sth & 0 & \cth
\end{bmatrix} \!, 
\hem \alpha = \pm1 ; &
\eeqn
\ese
where, 
$(\rho,\phi)$ are the polar coordinate variables of a constant $z$ plane.
Here $\alpha = 1$ and $\alpha = -1$ represent convergence and divergence respectively.
Since the above transformation is described by using a planar surface representation (the $z = 0$ plane) instead of a Gaussian reference sphere representation, the amplitude modifying factor $g$ appears here as $1/\sqrt{\cth}$ \cite{Bekshaev2010, ADNKVrt2020} instead of $\sqrt{\cth}$ \cite{RichardsWolf1959}.

Physically, the above transformation most commonly represents the curvature induced to a plane wave due to a lens, whose effective plane is considered as the $z = 0$ plane (just before the aperture). But generally, this transformation can also represent a system where an already existing curved wavefront beam may propagate and reach the $z = 0$ plane, where its amplitude would be described by Eqs. (\ref{R_operator}).


With the above understandings, we simplify Eqs. (\ref{R_operator}) and obtain the amplitude vector $\bcE'$ as
\bse \label{cEprime}
\beqn
& \bcE' = g \hpt \cE_0 \hehat 
= g \hpt \cE_0 \left( e_x \hxhat + e_y \hyhat + e_z \hzhat \right); & \\
& e_x = (\cth - 1) \cph\sph , & \\
& e_y = \cth\ssph + \ccph, & \\
& e_z = \alpha \hpt \sth\sph . & 
\eeqn
\ese
Clearly, $\bcE'$ becomes $\cE_0 \yhat$ as $\theta \rightarrow 0$, signifying that the effect of divergence or convergence diminishes as $Z_C \rightarrow \infty$. 


\subsection{Phase Variation}

The complete field $\bE'$ at the $z = 0$ plane just before the aperture is now obtained by including a phase term with the amplitude $\bcE'$. For convenience, we suppress the time dependent term $e^{-i\omega t}$, and include only a path dependent term. Without any loss of generality, we consider the zero reference phase $\Phi_\mathrm{ref} = 0$ at the aperture-center $O_A$. Then, as seen in Fig. \ref{Fig_Model}(a), a ray that has started from the position $O (\rho,\phi,0)$ is `yet to travel' an extra distance $\Delta r = O O_W$ before achieving phase-equality with the field at $O_A$; whereas, as seen in Fig. \ref{Fig_Model}(b), a ray that has reached the position $O (\rho,\phi,0)$ has `already travelled' an extra distance $\Delta r = O_W O$ after achieving phase-equality with the field at $O_A$. The phase term for the converging ($\alpha = 1$) and diverging ($\alpha = -1$) cases can then be expressed as $e^{-i \alpha \Delta\Phi}$, where
\beqn \label{DPhi_def}
& \Delta\Phi = n k \Delta r, \hem \Delta r = \sqrt{\rho^2 + Z_C^2} - Z_C, &
\eeqn
where, $k = 2\pi/\lambda$ ($\lambda = $ wavelength in free space), 
and $n$ is the refractive index of the medium in the $z > 0$ region for the chosen $\lambda$.
The complete field just before the aperture is then expressed as
\begin{\eq} \label{Eprime}
\bE' = \bcE' e^{-i \alpha \Delta\Phi} = g \hpt \cE_0 \hpt e^{-i \alpha \Delta\Phi} \hehat.
\end{\eq}

\subsection{Aperture Transmittance} \label{Subsec_Circ}

Subsequently, the field $\bE'$ passes through the aperture. The transmittance function of the circular aperture is defined by the circle function \cite{Goodman}
\begin{\eq} \label{circ_def}
\circf(u) = \left\{\begin{array}{lll}
1 & \mbox{for} & u < 1, \\
\frac{1}{2} & \mbox{for} & u = 1, \\
0 & \mbox{for} & u > 1, 
\end{array}
\right.
\end{\eq}
where, $u = \rho/a$ is a dimensionless radial variable, normalized \wrt the radius $a$ of the aperture. By applying this transmittance to the field $\bE'$, we obtain the aperture output field at the $z = 0$ plane as
\begin{\eq} \label{EA}
\bE_A = \circf(\rho/a) \hpt \bE' = \circf(\rho/a) \hpt g \hpt \cE_0 \hpt e^{-i \alpha \Delta\Phi} \hehat .
\end{\eq}

\subsection{Free Propagation}

Finally, the beam field propagates to a detector screen placed at a $z = D > 0$ plane (not shown in Fig. \ref{Fig_Model}). This propagation occurs through the linear homogeneous isotropic dielectric medium of refractive index $n$ in the $z > 0$ region. To understand the field transformation due to this propagation \cite{Goodman, NanoOpticsBook, Voelz}, we first determine the Fourier spectrum of $\bE_A$ [Eq. (\ref{EA})] as
\begin{\eq}
\tbE_A (k_x, k_y) = \dfrac{1}{4\pi^2} \int_{-\infty}^{\infty} \int_{-\infty}^{\infty}
\bE_A (x,y) \hpt e^{-i(k_x x + k_y y)} dx dy ,
\end{\eq}
where, $k_x \xhat$ and $k_y \yhat$ are the transverse components of a wavevector $\bk = n k \hpt \hat{\bk}$ that represents a constituent plane wave.
As the beam propagates to the $z = D$ plane, the spectrum $\tbE_A$ is multiplied by the transfer function 
\begin{\eq}
H_D (k_x, k_y) = e^{i k_z D}, \hem k_z = \left[ n^2 k^2 - (k_x^2 + k_y^2) \right]^{\frac{1}{2}},
\end{\eq}
and we obtain the Fourier spectrum of the field at $z = D$ as
\begin{\eq}
\tbE_D (k_x, k_y) = H_D (k_x, k_y) \hpt \tbE_A (k_x, k_y) .
\end{\eq}
Then, the final field at $z = D$ is obtained by inverse transforming $\tbE_D$ as
\begin{\eq} \label{ED}
\bE_D (x, y) = \int_{-\infty}^{\infty} \int_{-\infty}^{\infty}
\tbE_D (k_x, k_y) \hpt e^{i(k_x x + k_y y)} d k_x d k_y .
\end{\eq}

The above free propagation formulation is applicable to all distances $D$. But significant diffraction effects are observed only for $D \gg a$, which represents a typical experimental scenario. It is extremely difficult to find an analytical final expression for the above field $\bE_D$. So, in the following section, we discuss results by simulating the optical system.


\section{Simulated Field Properties} \label{Sec_Sim}


Considering the requirement of having a simple enough initial condition, 
we choose to define the amplitude function $\cE_0$ in an ideal Gaussian form as
\begin{\eq} \label{cE0Gauss}
\cE_0 = \cE_{00} \hpt e^{-\rho^2/w_0^2},
\end{\eq}
where, $\cE_{00}$ is the central magnitude, and $w_0$ is the effective Gaussian half-width. 
It is also understood from Eqs. (\ref{cEprime}) that all the relevant fields have nonzero longitudinal ($\zhat$) components. However, we make our final observations at a detector screen placed at a distance $D \gg a$, and the observation area of interest is considered around the beam axis with a length scale less than $a$. So, for our purpose, it is sufficient to explore only the transverse ($\xhat$ and $\yhat$) components of all the fields.

\subsection{Computational Considerations}

As understood from Fig. \ref{Fig_Model}, the maximum possible $\theta$ that gives a non-zero aperture transmittance is $\theta_{\mathrm{max}} = \atan(a/Z_C)$. To study a large enough divergence or convergence, we must then consider $Z_C$ values $Z_C \lesssim a$. If we consider typical values of $a \sim 10^{-3}$ m, then the oscillation of $e^{-i\alpha \Delta\Phi}$ [Eq. (\ref{DPhi_def})] at the aperture becomes too large to be appropriately sampled for fast Fourier transform (FFT) in a regular computer \cite{Voelz}. To avoid this problem, we scale down the size of $a$ to $\sim 10^{-4}$ m, and assign $Z_C$ values accordingly. This does not give any physically unacceptable result because we maintain the ranges $a \gg \lambda/n$ and $Z_C \gg \lambda/n$. We also consider the free propagation path length $D$ shorter than what is used in an actual experiment, but we maintain the range $D \gg a$. One may thus consider a real experimental scenario as a length-upscaled version of this simulated model.
Of course, the above problem does not arise for small divergence or convergence, when $Z_C \gg a$. But for consistency we use fixed $a$ and $D$ values, and corresponding appropriate $Z_C$ values, for both large and small divergences and convergences.

In the simulation we have observed results for many different parameter values. While the exact form of the $\bE_D$ field profile [Eq. (\ref{ED})] is different for each different set of parameter values, some significant general characteristics are consistently observed, which also agree with the subsequent experimental results [Sec. \ref{Sec_Exp}]. 
In the following subsections we demonstrate these essential features by considering the simulation parameters $\lambda = 632.8$ nm, $n = 1$, $w_0 = 2$ mm, $a = 0.6$ mm and $D = 160$ mm.

\subsection{Variation of \textit{Z}$\hspace{-0.08em}_C$} \label{Subsec_ZCvariation}

\begin{figure}
\includegraphics[width = \linewidth]{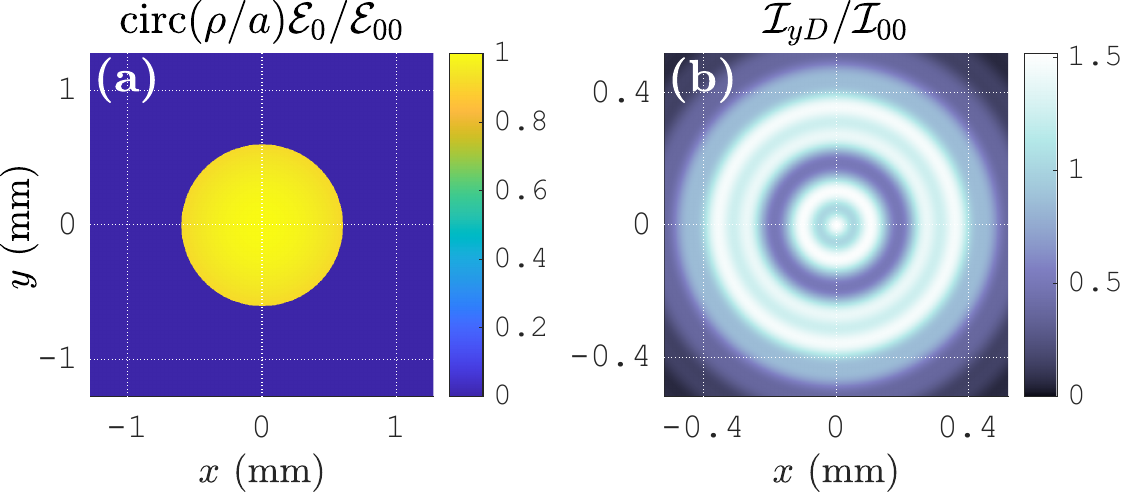}
\caption{
(a) Simulated field amplitude profile $\circf(\rho/a) \hpt \cE_0$ for $Z_C \rightarrow \infty$, considering the parameter values $w_0 = 2$ mm and $a = 0.6$ mm. (b) Corresponding intensity profile $\cI_{y D} \propto |\bE_D|^2$ at $z = D$, considering the parameter values $\lambda = 632.8$ nm, $n = 1$ and $D = 160$ mm. The $\cI_{y D}$ profile is normalized \wrt the initial maximum intensity $\cI_{00} \propto \cE_{00}^2$, revealing that the $\cI_{y D}$ values at some of the bright fringes are higher than $\cI_{00}$.
This happens because, as the dark regions in the diffraction pattern are created, the energy of the beam is redistributed in the bright regions.
}
{\color{grey}{\rule{\linewidth}{1pt}}}
\label{Fig_EAIyDz00}
\end{figure}

For $Z_C \rightarrow \infty$, the field $\bE_A$ [Eq. (\ref{EA})] becomes $\circf(\rho/a) \hpt \cE_0 \yhat$. 
The corresponding free propagated field $\bE_D$, as obtained via Eq. (\ref{ED}), is thus only $\yhat$ polarized.
The $\circf(\rho/a) \hpt \cE_0$ amplitude profile and the simulated intensity profile $\cI_{y D} \propto |\bE_D|^2$ are shown in Fig. \ref{Fig_EAIyDz00}. Characteristic circular aperture diffraction rings are observed in the $\cI_{y D}$ profile. This intensity pattern is not an ideal Airy pattern, because the considered propagation distance $D = 160$ mm is too short to validate Fraunhofer approximation for the aperture radius $a = 0.6$ mm \cite{Goodman}. However, we have observed in the simulation that, as the distance is increased, the pattern of Fig. \ref{Fig_EAIyDz00}(b) gradually transforms to an ideal Airy pattern.

To obtain a nonzero $\xhat$ component in $\bE_D$, we require a nonzero $\xhat$ component in $\bE_A$.
This is achieved when finite values of $Z_C$ are considered. The transverse amplitude of $\bE_A$ can be expressed by using Eq. (\ref{EA}) and excluding the $e^{-i \alpha \Delta\Phi}$ term as
\bse \label{EperpA_ExyA}
\beqn
& \bcE_{\perp A} = \cE_{xA} \hxhat + \cE_{yA} \hyhat; \hem &\\
& \cE_{xA} = \cE_A e_x, \hem \cE_{yA} = \cE_A e_y, \hem \cE_A = \circf(\rho/a) \hpt g \hpt \cE_0 . \hem\hem &
\eeqn
\ese
The above expression of $\bcE_{\perp A}$ is valid for both the diverging and the converging cases [Fig. \ref{Fig_Model}], as understood from Eqs. (\ref{cEprime}). 
However, the phase terms $e^{-i \Delta\Phi}$ for convergence and $e^{+i \Delta\Phi}$ for divergence make the final $\bE_D$ field substantially different. To understand the difference, one can form a qualitative visualization based on a geometrical optics picture as follows: in the converging case, if $Z_C > D$, the beam converges throughout the entire distance $D$ to reach the detector; and if $Z_C < D$, the beam first converges through a distance $Z_C$ (to the point $O_C$ in Fig. \ref{Fig_Model}(a)) and then diverges through a distance $D - Z_C$ to reach the detector. On the contrary, in the diverging case, the beam diverges throughout the entire distance $D$ for all $Z_C$ values. Hence, the overall beam power reaching the detector (because of its fixed finite area) is larger in the converging case as compared to that in the diverging case, and this difference can be made quite significant by choosing appropriate values of $Z_C$.

\begin{figure}
\includegraphics[width = \linewidth]{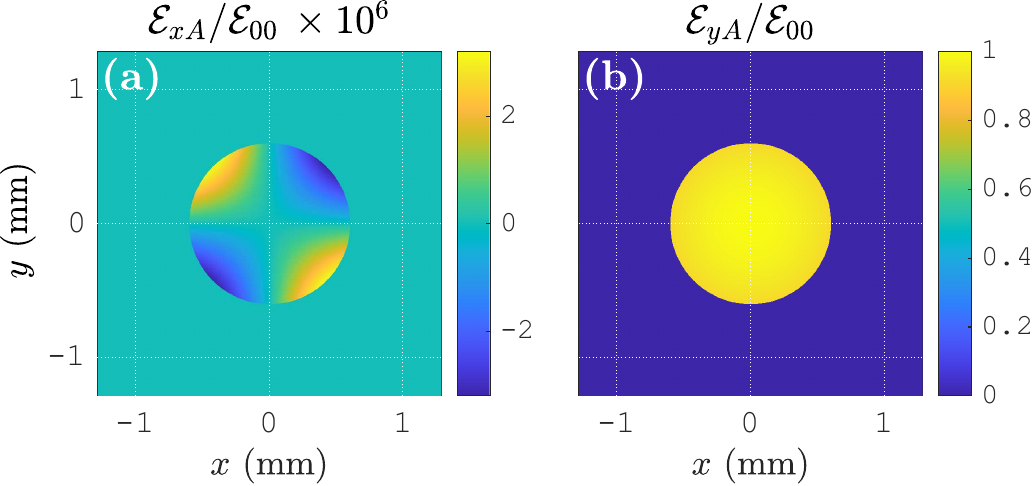}
\caption{
Simulated profiles of the components $\cE_{xA}$ and $\cE_{yA}$ of the transverse field amplitude $\bcE_{\perp A}$, considering $Z_C = D = 160$ mm. 
The $\cE_{xA}$ values in (a) are several orders of magnitude less than $\cE_{00}$, and hence is represented here by multiplying with a factor $10^6$. This convention is also followed in the subsequent figures, wherever applicable. 
The $\cE_{yA}$ profile in (b) is approximately equal to the $\circf(\rho/a) \hpt \cE_0$ profile of Fig. \ref{Fig_EAIyDz00}(a), as $\theta_\mathrm{max} \ll 1$.
For both converging ($\alpha = 1$) and diverging ($\alpha = -1$) cases, the same transverse amplitude $\bcE_{\perp A}$ appears for a given $Z_C$, but the complete transverse fields differ due to the phase term $e^{-i \alpha \Delta\Phi}$.
}
{\color{grey}{\rule{\linewidth}{1pt}}}
\label{Fig_ExyA_lowConv}
\end{figure}

\begin{figure*}
\includegraphics[width = \linewidth]{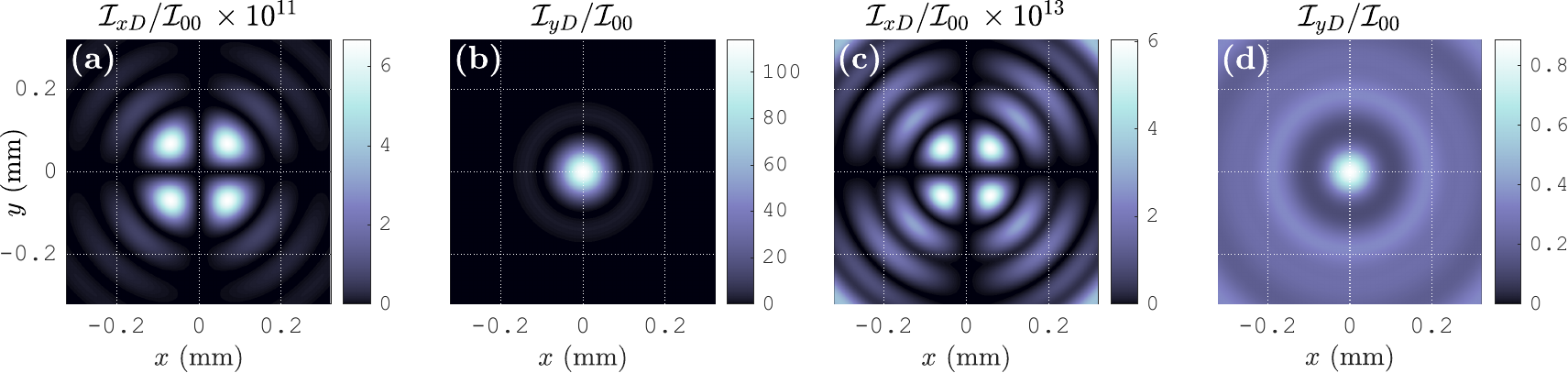}
\caption{
Simulated intensity profiles $\cI_{xD}$ and $\cI_{yD}$ for [(a),(b)] convergence ($\alpha = 1$) and [(c),(d)] divergence ($\alpha = -1$), considering $Z_C = D = 160$ mm.
}
{\color{grey}{\rule{\linewidth}{1pt}}}
\label{Fig_IxyD_lowDivConv}
\end{figure*}

To give an example, we consider $Z_C = D = 160$ mm, which gives $\theta_{\mathrm{max}} = \atan(a/Z_C) \approx 0.215\deg$ for $a = 0.6$ mm. The corresponding $\cE_{xA}$ and $\cE_{yA}$ profiles [Eqs. (\ref{EperpA_ExyA})] are shown in Fig. \ref{Fig_ExyA_lowConv} (same profiles for both convergence and divergence).
The transverse component of $\bE_D$, now with a nonzero $\xhat$ component for a finite $Z_C$, takes the form
\begin{\eq} \label{EperpD}
\bE_{\perp D} = E_{xD} \hxhat + E_{yD} \hyhat ,
\end{\eq}
with a total intensity
\begin{\eq} \label{IperpD_IxD_IyD}
\cI_{\perp D} = \cI_{xD} + \cI_{yD} , \hemm 
\cI_{xD} \propto |E_{xD}|^2, \hem \cI_{yD} \propto |E_{yD}|^2.
\end{\eq}
The simulated intensity profiles $\cI_{xD}$ and $\cI_{yD}$, 
for both convergence and divergence, are shown in Fig. \ref{Fig_IxyD_lowDivConv}.
Two remarkable observations can be made from these profiles:

\benum

\item Nonzero $\cI_{xD}$ appears for both convergence and divergence, but the maximum $\cI_{xD}$ value in the converging case [Fig. \ref{Fig_IxyD_lowDivConv}(a)] is substantially higher than the corresponding value in the diverging case [Fig. \ref{Fig_IxyD_lowDivConv}(c)].

\item The maximum $\cI_{yD}$ value, as compared to that in Fig. \ref{Fig_EAIyDz00}(b), is reduced in the diverging case [Fig. \ref{Fig_IxyD_lowDivConv}(d)], and substantially increased in the converging case [Fig. \ref{Fig_IxyD_lowDivConv}(b)].

\eenum

The above observations thus confirm the achievement of a nonzero $E_{xD} \xhat$ field, as well as the increase in the detected beam power due to convergence.


However, the above differences are significant only for $Z_C \sim D$. 
To obtain sufficiently high convergence and divergence, we choose $Z_C \lesssim a$ (hence $Z_C \ll D$). Because of this choice, the converging beam eventually diverges through a distance $D - Z_C$ which is comparable to the diverging distance $D$ in the diverging case. This makes the final $\bE_{\perp D}$ field comparable in both cases, though the beam power in the converging case maintains a value higher than that in the diverging case.

\begin{figure}
\includegraphics[width = \linewidth]{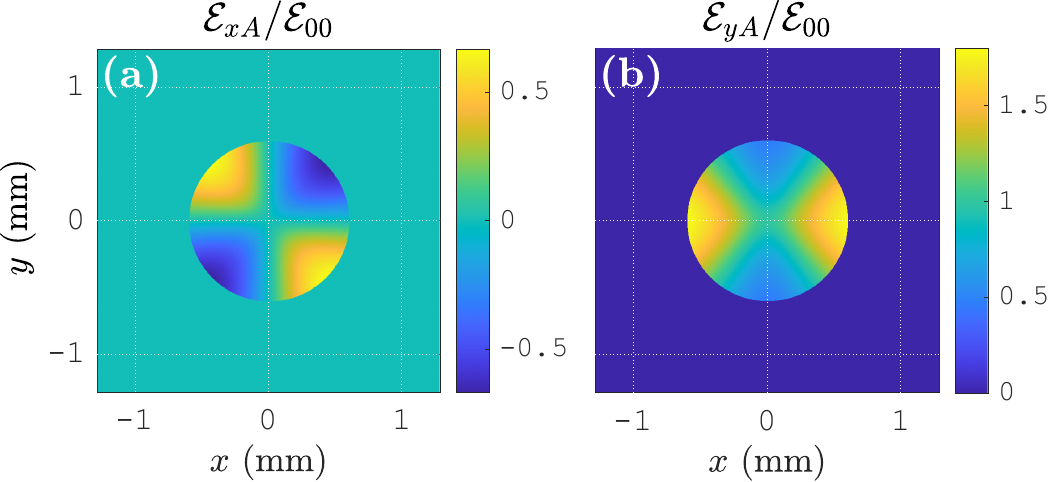}
\caption{
Simulated $\cE_{xA}$ and $\cE_{yA}$ profiles, considering $Z_C = 0.16$ mm (same for both $\alpha = \pm1$). 
Some regions are observed in (b) where the $\cE_{yA}$ values are higher than the initial maximum magnitude $\cE_{00}$. This happens because the factor $g = 1/\sqrt{\cos\theta}$ [Eqs. (\ref{R_operator})] locally squeezes the wavefront surface elements in the planar representation of the curved wavefront transformation \cite{Bekshaev2010, ADNKVrt2020}, thus increasing the local intensity and the field magnitude.
}
{\color{grey}{\rule{\linewidth}{1pt}}}
\label{Fig_ExyA}
\end{figure}

\begin{figure*}
\includegraphics[width = \linewidth]{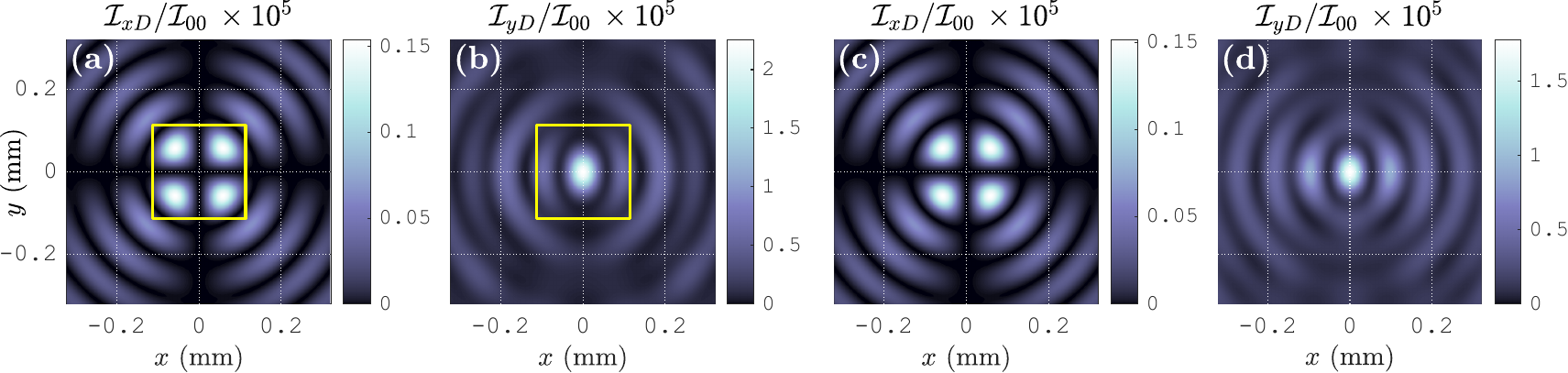}
\caption{
Simulated intensity profiles $\cI_{xD}$ and $\cI_{yD}$ for [(a),(b)] convergence ($\alpha = 1$) and [(c),(d)] divergence ($\alpha = -1$), considering $Z_C = 0.16$ mm.
The central region of interest, where the polarization characteristics are to be studied, is marked in (a) and (b).
}
{\color{grey}{\rule{\linewidth}{1pt}}}
\label{Fig_IxyD}
\end{figure*}

To give an example, we consider $Z_C = 0.16$ mm, which gives $\theta_{\mathrm{max}} \approx 75\deg$ for $a = 0.6$ mm.
The resulting $\cE_{xA}$ and $\cE_{yA}$ profiles are shown in Fig. \ref{Fig_ExyA}. 
The corresponding intensity profiles $\cI_{xD}$ and $\cI_{yD}$, for both convergence and divergence, are shown in Fig. \ref{Fig_IxyD}.
As opposed to the fringes seen in Figs. \ref{Fig_EAIyDz00}(b) and \ref{Fig_IxyD_lowDivConv}(d), the new $\cI_{yD}$ fringes of Figs. \ref{Fig_IxyD}(b) and \ref{Fig_IxyD}(d) are distorted away from a uniform circular nature due to high convergence and divergence. We have observed in the simulation that, in the vicinity of the considered $Z_C = 0.16$ mm, these fringes vary significantly with $Z_C$ variations even as small as $0.01$ mm.
In contrast, the fringes of the $\cI_{xD}$ profiles [Figs. \ref{Fig_IxyD}(a), \ref{Fig_IxyD}(c)] remain approximately the same for small $Z_C$ variations in the vicinity of a given $Z_C$.

\subsection{Polarization Characteristics} \label{Subsec_PolarizationSim}

The field $\bE_{\perp D}$ [Eq. (\ref{EperpD})] is significantly complicated, and hence its true nature is not entirely revealed by only the intensity profiles of Figs. \ref{Fig_IxyD_lowDivConv} and \ref{Fig_IxyD}. It is necessary to explore the polarization characteristics to fully understand the nature of $\bE_{\perp D}$. For this purpose, we aim to study the polarizations in the most intense central region of the beam field. We consider the converging case with $Z_C = 0.16$ mm as an example. 
The central region of interest is marked in Figs. \ref{Fig_IxyD}(a) and \ref{Fig_IxyD}(b).

\begin{figure*}
\includegraphics[width = \linewidth]{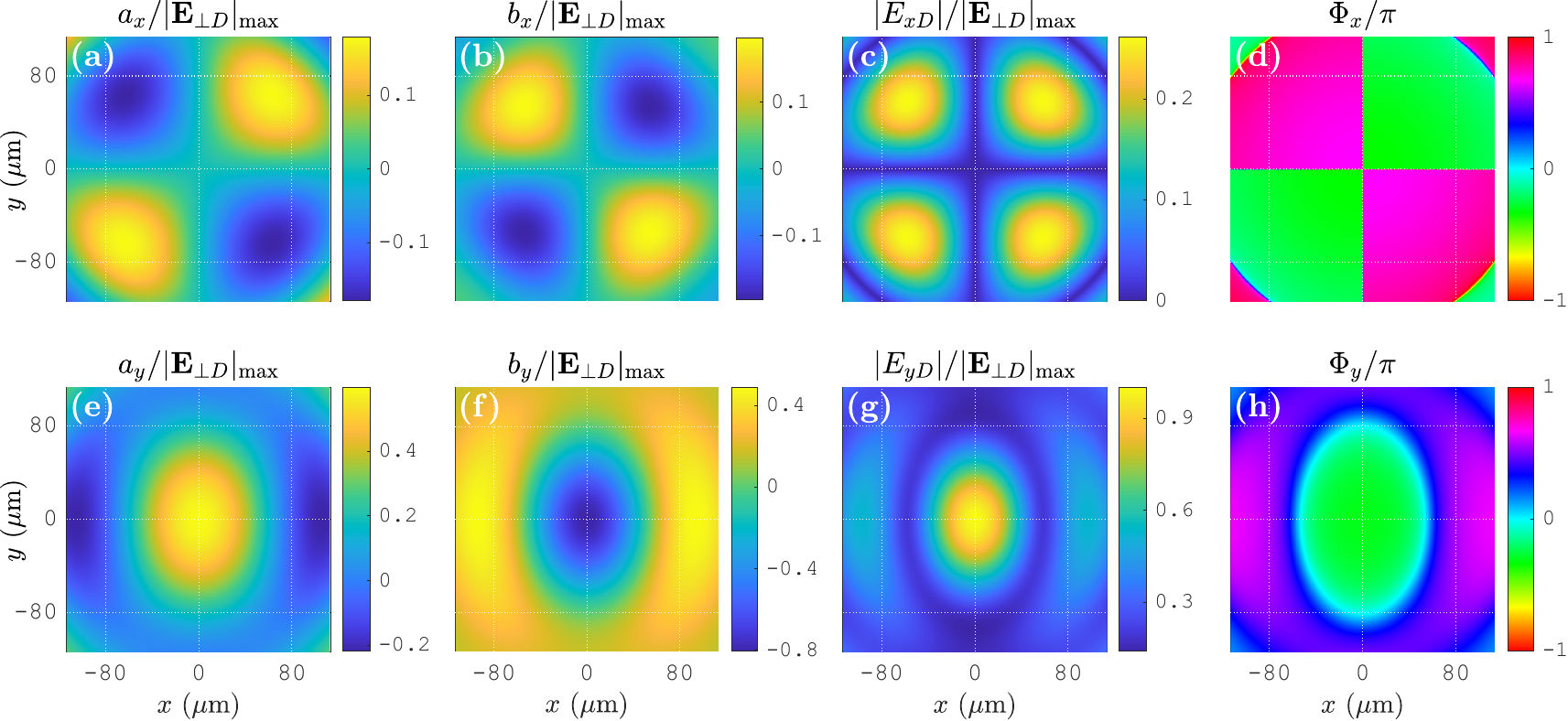}
\caption{
Simulated $a_q$, $b_q$, $|E_{qD}|$ and $\Phi_q$ profiles ($q = x,y$), considering $Z_C = 0.16$ mm and $\alpha = 1$. 
The $a_q$, $b_q$ and $|E_{qD}|$ profiles are normalized \wrt $|\bE_{\perp D}|_\mathrm{max}$, i.e. the maximum magnitude of $\bE_{\perp D}$ in the region of interest.
}
{\color{grey}{\rule{\linewidth}{1pt}}}
\label{Fig_abEPhi}
\end{figure*}

The functions $E_{xD}$ and $E_{yD}$ [Eq. (\ref{EperpD})] are complex amplitude functions of the form
\bse \label{Exy_abxy_full}
\beqn
& E_{qD} = a_q + i b_q = |E_{qD}| \hpt e^{i \Phi_q}, \hem (q = x, y), & \\
& |E_{qD}| = \sqrt{a_q^2 + b_q^2} \hpt, \hem \Phi_q = \atan(b_q/a_q), &
\eeqn
\ese
where, $|E_{qD}|$ and $\Phi_q$ are respectively the field magnitude and phase functions. To consider the $\Phi_q$ values in a full $(-\pi,\pi]$ range, the following convention for the evaluation of the inverse tangent function is followed:
\bse \label{PhaseConvention}
\beqn
0 \leq \Phi_q \leq \pi/2 & \hem \mbox{for} \hem & a_q \geq 0, b_q \geq 0, \\
\pi/2 < \Phi_q \leq \pi & \hem \mbox{for} \hem & a_q < 0, b_q \geq 0, \\
-\pi < \Phi_q < -\pi/2 & \hem \mbox{for} \hem & a_q < 0, b_q < 0, \\
-\pi/2 \leq \Phi_q < 0 & \hem \mbox{for} \hem & a_q \geq 0, b_q < 0. 
\eeqn
\ese
The simulated $a_q$, $b_q$, $|E_{qD}|$ and $\Phi_q$ profiles ($q = x,y$) for the presently considered parameters are shown in Fig. \ref{Fig_abEPhi}.

The $\Phi_x$ profile of Fig. \ref{Fig_abEPhi}(d) shows $\pi$ phase jumps along the $x$ and $y$ axes, which must be interpreted carefully. The profiles of Figs. \ref{Fig_abEPhi}(a) and \ref{Fig_abEPhi}(b) show that 
the following conditions are satisfied in the vicinity of the $x$ and $y$ axes in the region of interest:
\benum
\item $(a_x \geq 0 \mbox{ AND } b_x \leq 0)$ for 
$$ (x \geq 0 \mbox{ AND } y \geq 0) \mbox{ OR } 
(x \leq 0 \mbox{ AND } y \leq 0), $$

\item $(a_x < 0 \mbox{ AND } b_x > 0)$ for 
$$ (x < 0 \mbox{ AND } y > 0) \mbox{ OR } 
(x > 0 \mbox{ AND } y < 0). $$
\eenum
So, at a given time $t$, if $E_{xD} \xhat$ is directed along $+\xhat$ in the first and third quadrants (within the region of interest), then at the same time it is directed along $-\xhat$ in the second and fourth quadrants. This sign flip information is of course not contained in the magnitude $|E_{xD}|$ [Fig. \ref{Fig_abEPhi}(c)]. So $\pi$ phase jumps appear in the $\Phi_x$ profile to contain the sign information, since $e^{\pm i\pi} = -1$. Thus, these phase jumps signify the preservation of mathematical consistency, but not physical wavefront dislocations \cite{ADNKVrt2020}.

\begin{figure}
\includegraphics[width = \linewidth]{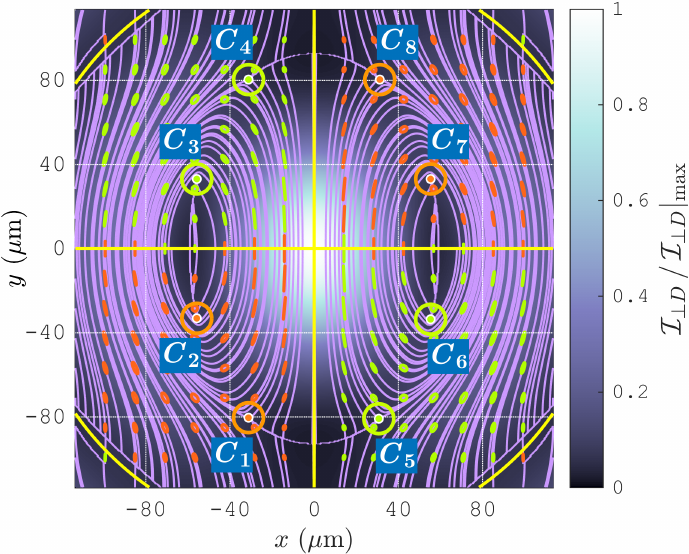}
\caption{
Simulated polarization profile of $\bE_{\perp D}$, considering $Z_C = 0.16$ mm and $\alpha = 1$. The right and left handed elliptical polarizations are shown respectively in dark orange and light green. The major axis orientations of the polarization ellipses are represented by purple streamlines. The $C$ point singularities $\{C_1,C_2,\cdots,C_8\}$ are marked. The $L$ line singularities are shown in yellow. The intensity profile $\cI_{\perp D} \propto |\bE_{\perp D}|^2$ is shown in the background, normalized with respect to its maximum value $\cI_{\perp D}|_\mathrm{max}$.
These conventions are also followed in the subsequent field profile figures as applicable. 
}
{\color{grey}{\rule{\linewidth}{1pt}}}
\label{Fig_EstreamSim}
\end{figure}

With the complete available field information we then plot the transverse polarization profile of the beam field, as shown in Fig. \ref{Fig_EstreamSim}. The orientation patterns of the major axes of the polarization ellipses are shown via purple streamlines. 
Some special features of this profile are recognized as polarization singularities \cite{ BerryHannay1977, TCN1978, Nye83a, Nye83b, Hajnal87a, Hajnal87b, NH1987, DH1994, Gbur}.
We identify the points $\{C_1,C_2,\cdots,C_8\}$ where the polarizations are purely circular, and hence the orientations are indeterminate. 
These points are $C$ point polarization singularities. 
In addition, the polarizations are right elliptical (REP) in some regions (polarizations shown in dark orange), and left elliptical (LEP) in some others (polarizations shown in light green). 
This implies that there exist line boundaries between the REP and LEP regions where the polarizations are linear, and hence the handedness is undefined. These line boundaries are $L$ line polarization singularities of the beam field, represented in yellow in Fig. \ref{Fig_EstreamSim}.

\subsection{\textit{C} Point Polarization Singularities} \label{Subsec_CpointSim}

\begin{figure*}
\includegraphics[width = \linewidth]{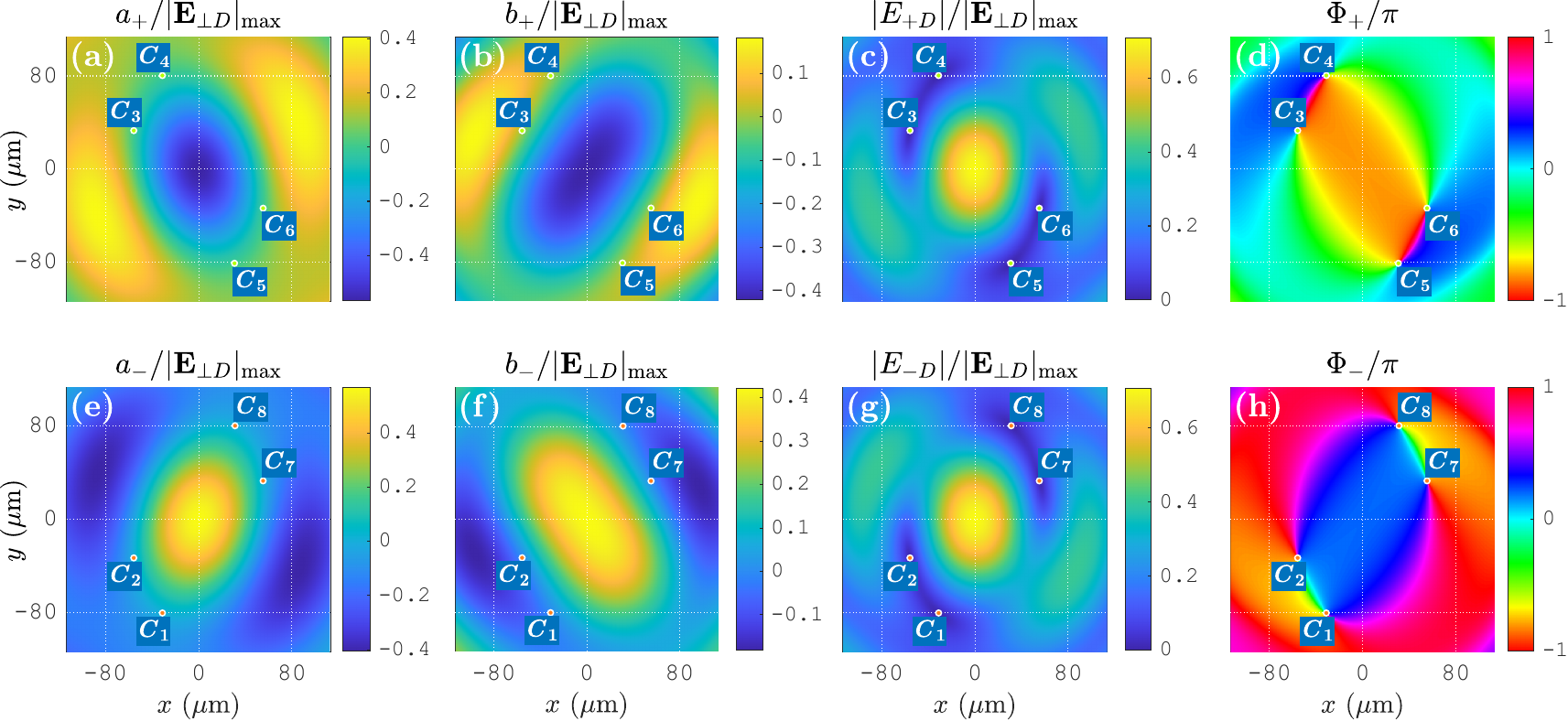}
\caption{
Simulated $a_q$, $b_q$, $|E_{qD}|$ and $\Phi_q$ profiles ($q = \pm$) corresponding to the $\bE_{\perp D}$ profile of Fig. \ref{Fig_EstreamSim}. The $C$ point singularity positions are marked.
}
{\color{grey}{\rule{\linewidth}{1pt}}}
\label{Fig_abEPhiPM}
\end{figure*}

To further explore the properties of the $C$ point singularities, we express $\bE_{\perp D}$ [Eq. (\ref{EperpD})] in terms of its constituent $\sighat^\pm$ spin polarizations. By substituting $\xhat = (\sighat^+ + \sighat^-)/\sqrt{2}$ and $\yhat = (\sighat^+ - \sighat^-)/\sqrt{2}i$ in Eq. (\ref{EperpD}), and rearranging, we obtain 
\bse \label{EperpD_sigmaPM}
\beqn
& \bE_{\perp D} = E_{+D} \hsighat^+ + E_{-D} \hsighat^-, & \\
& E_{\pm D} = (E_{xD} \mp i E_{yD})/\sqrt{2} \, . &
\eeqn
\ese
Here, $E_{\pm D}$ are the complex amplitudes of the constituent $\sighat^\pm$ polarizations, and can be expressed in the form of Eqs. (\ref{Exy_abxy_full}) with $q = \pm$, as follows:
\bse \label{EPM_abPM_full}
\beqn
& E_{\pm D} = a_\pm + i b_\pm = |E_{\pm D}| \hpt e^{i \Phi_\pm}, & \\
& |E_{\pm D}| = \sqrt{a_\pm^2 + b_\pm^2} \hpt, \hem \Phi_\pm = \atan(b_\pm/a_\pm), & \label{EPhPM}
\eeqn
\ese
where, the convention of Eqs. (\ref{PhaseConvention}) is followed to evaluate $\Phi_\pm$.
The simulated $a_\pm$, $b_\pm$, $|E_{\pm D}|$ and $\Phi_\pm$ profiles corresponding to the presently considered field $\bE_{\perp D}$ [Fig. \ref{Fig_EstreamSim}] are shown in Fig. \ref{Fig_abEPhiPM}.

As seen in Figs. \ref{Fig_abEPhiPM}(a) and \ref{Fig_abEPhiPM}(b), $a_+ = b_+ = |E_{+D}| = 0$ at $\{C_3, C_4, C_5, C_6\}$, which makes $\Phi_+$ indeterminate [Eq. (\ref{EPhPM})]. These points are thus the phase singularities of the $E_{+ D} \hsighat^+$ component field. Since $|E_{+D}|$ is zero, the total field $\bE_{\perp D}$ is purely $\sighat^-$ polarized at these points. However, $|E_{+D}|$ is non-zero at the surrounding points, which makes $\bE_{\perp D}$ elliptically polarized. This implies that $\{C_3, C_4, C_5, C_6\}$ are points of isolated $\sighat^-$ polarizations of $\bE_{\perp D}$. This explains why these points are $C$ point polarization singularities of $\bE_{\perp D}$ [Fig. \ref{Fig_EstreamSim}]. The formation of the $\sighat^+$ polarized $C$ point singularities at $\{C_1, C_2, C_7, C_8\}$ also can be explained likewise by analyzing the $a_-$, $b_-$, $|E_{-D}|$ and $\Phi_{-}$ profiles [Figs. \ref{Fig_abEPhiPM}(e)--\ref{Fig_abEPhiPM}(h)], and by identifying the fact that these points are the phase singularities of the $E_{- D} \hsighat^-$ component field.

By observing the senses of phase increase (clockwise or counterclockwise) around the singularities in the $\Phi_\pm$ profiles [Figs. \ref{Fig_abEPhiPM}(d) and \ref{Fig_abEPhiPM}(h)], the topological charges are obtained as follows:
\benum
\item At $C_1$ and $C_8$, the $\Phi_-$ phase singularities have topological charges $\ft = -1$.
\item At $C_2$ and $C_7$, the $\Phi_-$ phase singularities have topological charges $\ft = +1$.
\item At $C_3$ and $C_6$, the $\Phi_+$ phase singularities have topological charges $\ft = -1$.
\item At $C_4$ and $C_5$, the $\Phi_+$ phase singularities have topological charges $\ft = +1$.
\eenum
These have the following implications \cite{NKVMonstar, ADNKVBrew2021} in reference to the streamline patterns of Fig. \ref{Fig_EstreamSim}:
\benum
\item In the region of $C_1$ and $C_8$, a non-singular $\sighat^+$ polarized field and a $\sighat^-$ polarized field with $\ft = -1$ phase singularities are superposed. These superpositions create star patterns \cite{BerryHannay1977, Gbur} of the streamlines.
\item In the region of $C_2$ and $C_7$, a non-singular $\sighat^+$ polarized field and a $\sighat^-$ polarized field with $\ft = +1$ phase singularities are superposed. These superpositions create lemon patterns \cite{BerryHannay1977, Gbur} of the streamlines.
\item In the region of $C_3$ and $C_6$, a non-singular $\sighat^-$ polarized field and a $\sighat^+$ polarized field with $\ft = -1$ phase singularities are superposed. These superpositions create lemon patterns of the streamlines.
\item In the region of $C_4$ and $C_5$, a non-singular $\sighat^-$ polarized field and a $\sighat^+$ polarized field with $\ft = +1$ phase singularities are superposed. These superpositions create star patterns of the streamlines.
\eenum
The singularities in the streamline patterns of the $\bE_{\perp D}$ field profile [Fig. \ref{Fig_EstreamSim}] are thus completely explained by the above observations.

\subsection{Experimentally Observable Quantities} \label{Subsec_ExpObservable}

The functions $a_q$, $b_q$ and $\Phi_q$ ($q = x, y, +, -$) are not straightforwardly observable without implementing specially designed phase measurement methods \cite{FFT1982, Zhao2017, ADNKVSPIE2023}. Instead, in a typical experiment, one observes the intensity profiles
\beqn
& \cI_{\perp D} \propto |\bE_{\perp D}|^2, \hem \cI_{xD} \propto |E_{xD}|^2, \hem \cI_{yD} \propto |E_{yD}|^2, & \nonumber \\
& \cI_{\pm D} \propto |E_{\pm D}|^2, \hem \cI_{d^\pm D} \propto |\dhat^\pm\cdot\bE_{\perp D}|^2, & \nonumber
\eeqn
where, $\dhat^\pm = (\xhat \pm \yhat)/\sqrt{2}$ represent the $\pm 45\deg$ linear polarizations.
Clearly, $\cI_{\perp D}$ is the directly observed intensity; whereas, the $\cI_{xD}$, $\cI_{yD}$ and $\cI_{d^\pm D}$ profiles are observed by passing the $\bE_{\perp D}$ field through an appropriately oriented polarizer. To observe the $\cI_{\pm D}$ profiles, the $\bE_{\perp D}$ field is first passed through a quarter wave plate (QWP) with its fast axis oriented along $\yhat$. Then, by passing the transformed field through a polarizer oriented along $\dhat^\pm$, the $\cI_{\pm D}$ intensity profiles are extracted and observed. These observations lead to the determination of the Stokes parameter profiles \cite{Goldstein}
\beqn
& S_0 \propto \cI_{\perp D}, \hem S_1 \propto \cI_{xD} - \cI_{yD}, & \nonumber \\
& S_2 \propto \cI_{d^+ D} - \cI_{d^- D}, \hem S_3 \propto \cI_{+ D} - \cI_{- D}, & \nonumber
\eeqn
and the normalized Stokes parameter profiles
\begin{\eq}
s_j = S_j/S_0, \hem j = 0,1,2,3.
\end{\eq}
Using these parameters, the polarization and streamline patterns of Fig. \ref{Fig_EstreamSim} can be completely reproduced.

\begin{figure*}
\includegraphics[width = \linewidth]{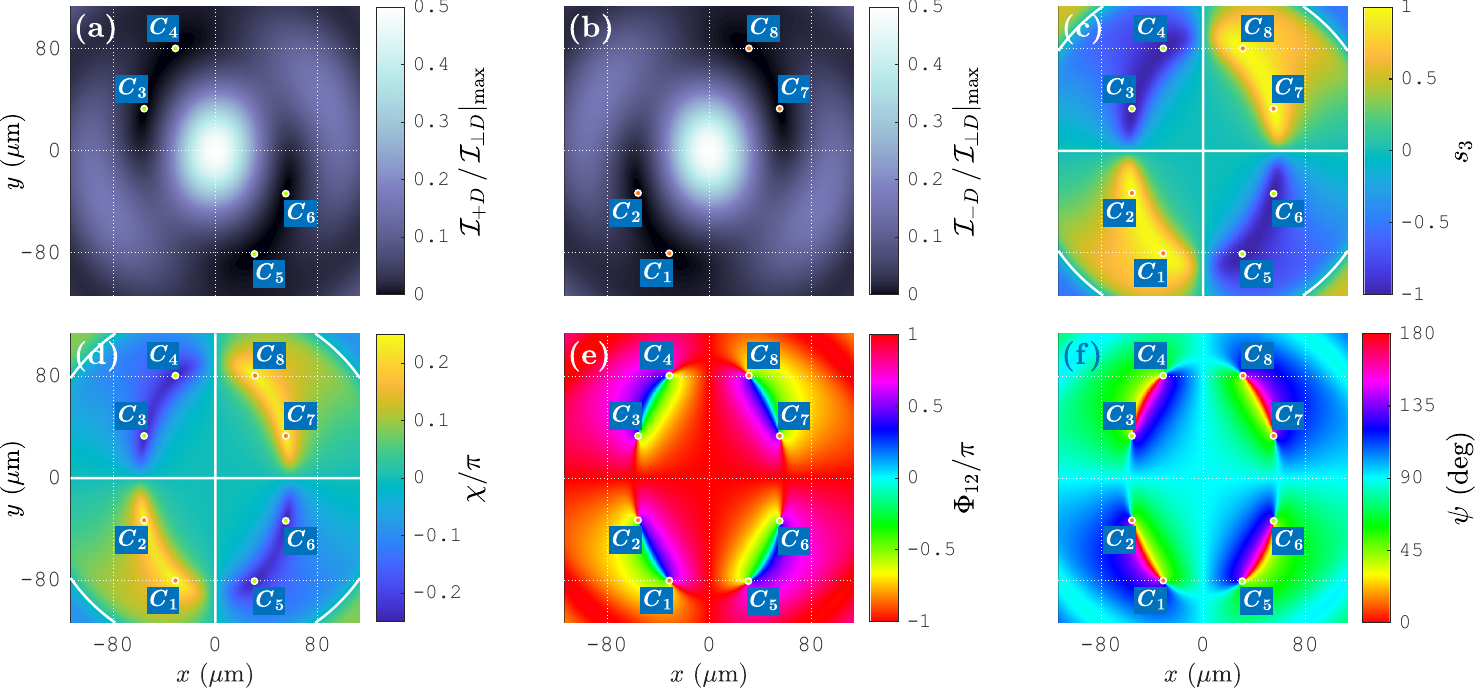}
\caption{
Simulated profiles of intensities $\cI_{\pm D}$, normalized Stokes parameter $s_3$, ellipticity $\chi$ and major axis orientation $\psi$ of the polarization ellipses, and phase $\Phi_{12}$ of the complex Stokes parameter $S_1 + i S_2$, corresponding to the $\bE_{\perp D}$ profile of Fig. \ref{Fig_EstreamSim}. The $C$ point singularity positions are marked.
At the $L$ line singularities we get $s_3 = \chi = 0$, which are marked in (c) and (d).
}
{\color{grey}{\rule{\linewidth}{1pt}}}
\label{Fig_StokesQuantitiesSim}
\end{figure*}

In Fig. \ref{Fig_StokesQuantitiesSim} we show some simulated function profiles which are experimentally obtainable, and are particularly relevant to the polarization singularities. The intensity profiles $\cI_{+ D}$ [Fig. \ref{Fig_StokesQuantitiesSim}(a)] and $\cI_{- D}$ [Fig. \ref{Fig_StokesQuantitiesSim}(b)] show minima at the appropriate singularity points. The normalized Stokes parameter $s_3$ [Fig. \ref{Fig_StokesQuantitiesSim}(c)] takes the value $+1$ at $\{C_1,C_2,C_7,C_8\}$ signifying $\sighat^+$ polarizations, and the value $-1$ at $\{C_3,C_4,C_5,C_6\}$ signifying $\sighat^-$ polarizations. Correspondingly, the ellipticity \cite{Goldstein}
\begin{\eq}
\chi = \dfrac{1}{2} \asin s_3
\end{\eq}
takes the values $+\pi/4$ at $\{C_1,C_2,C_7,C_8\}$ and $-\pi/4$ at $\{C_3,C_4,C_5,C_6\}$ [Fig. \ref{Fig_StokesQuantitiesSim}(d)]. We also get $s_3 = \chi = 0$ at the $L$ line singularities, which signify linear polarizations.
The phase $\Phi_{12}$ of the complex Stokes parameter $S_1 + i S_2$, determined by considering the convention of Eqs. (\ref{PhaseConvention}), represent the phase difference \cite{BliokhRev2019}
\begin{\eq} \label{Phi12}
\Phi_{12} = \Phi_- - \Phi_+.
\end{\eq}
The $\Phi_{12}$ profile of Fig. \ref{Fig_StokesQuantitiesSim}(e) is thus a combined representation of the individual phase profiles $\Phi_{+}$ [Fig. \ref{Fig_abEPhiPM}(d)] and $\Phi_{-}$ [Fig. \ref{Fig_abEPhiPM}(h)]. Finally, the ellipse orientation $\psi$, defined as \cite{Goldstein}
\begin{\eq}
\psi = \Phi_{12}/2 \, ,
\end{\eq}
and re-expressed in the range $[0\deg,180\deg)$ for convenience, is shown in Fig. \ref{Fig_StokesQuantitiesSim}(f). It is easily seen that this $\psi$ variation in the beam field completely describes the local orientations of the streamlines in Fig. \ref{Fig_EstreamSim}.

\subsection{Additional Examples} \label{Subsec_AddExampleSim}

\begin{figure*}
\includegraphics[width = \linewidth]{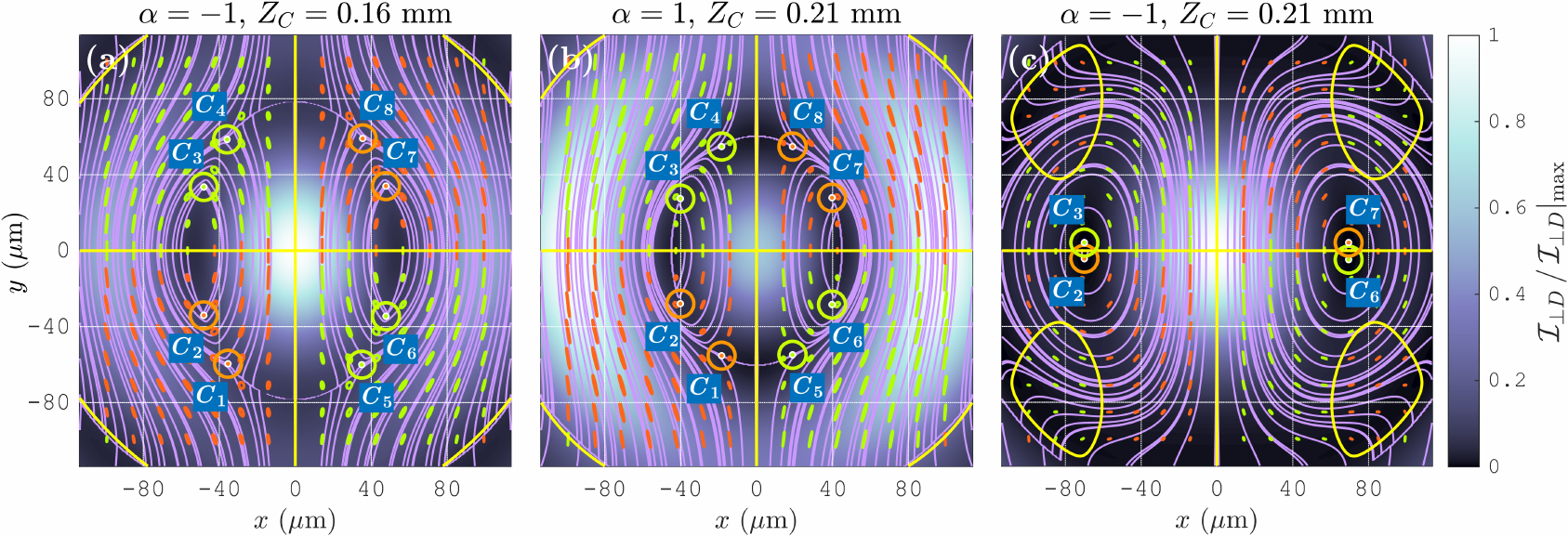}
\caption{
Simulated $\bE_{\perp D}$ profiles for some additional example cases. 
The $C$ points similar in nature to those in Fig. \ref{Fig_EstreamSim} are identified by the same names. It is to be noticed that not all similar $C$ points appear in all of the profiles.
}
{\color{grey}{\rule{\linewidth}{1pt}}}
\label{Fig_EStreamSimExtra}
\end{figure*}

In Sec. \ref{Subsec_PolarizationSim}--\ref{Subsec_ExpObservable} we have demonstrated the field characteristics by considering the example of the converging case with $Z_C = 0.16$ mm. As $Z_C$ is changed, the exact $\bE_{\perp D}$ profile varies, but its general properties remain similar.
For example, in a given area of interest, the number of $C$ points can be different for different $Z_C$ values. But the origin of each $C$ point can be understood in a similar way as described in Sec. \ref{Subsec_CpointSim}. In addition, various REP and LEP regions can appear in the given area. But their boundaries always become $L$ line singularities. 
A few more example field profiles in this regard are shown in Fig. \ref{Fig_EStreamSimExtra}.

Nevertheless, these effects are observed only when $E_{xD}$ and $E_{yD}$ [Eq. (\ref{EperpD})] are comparable. For very low divergence, the $E_{yD} \yhat$ component significantly dominates, and hence the field $\bE_{\perp D}$ remains approximately $\yhat$ polarized everywhere.


\section{Experimental Observations} \label{Sec_Exp}


\subsection{Experimental Setup}

\begin{figure}
\includegraphics[width = \linewidth]{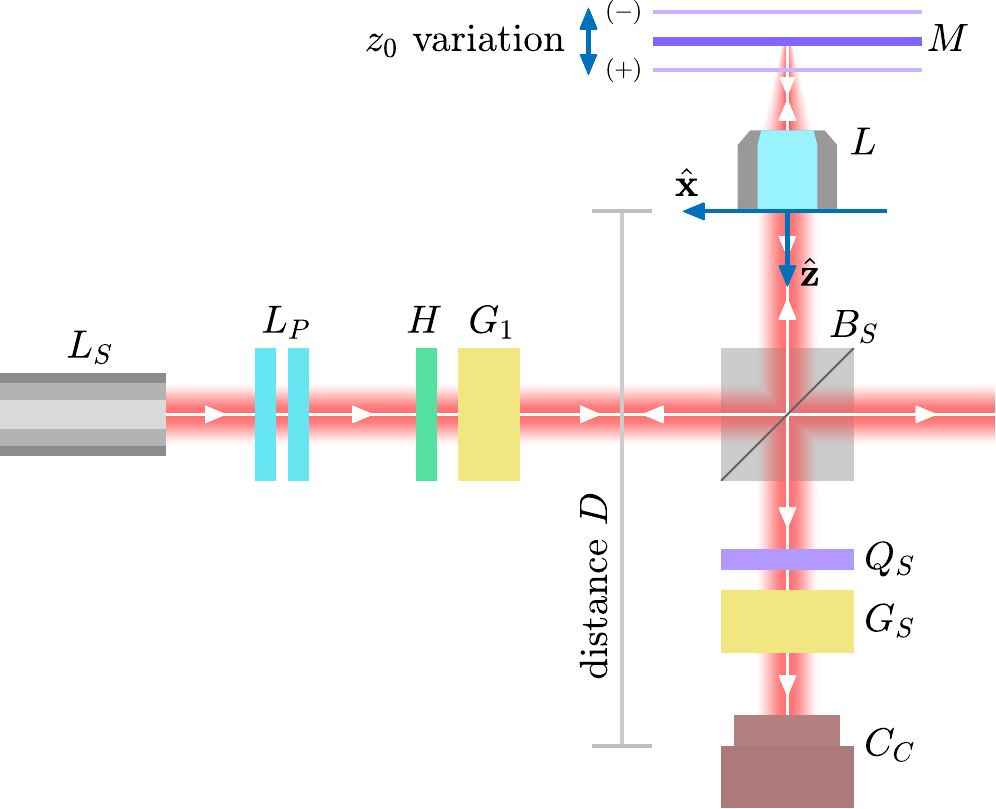}
\caption{The experimental setup (description in the text).
}
{\color{grey}{\rule{\linewidth}{1pt}}}
\label{Fig_ExpSetup}
\end{figure}

For the purpose of experimental observations, we build an optical system based on a normal incidence and reflection scheme \cite{NovotnyFocusImage2001, PetrovFocusedBeamRT2005, NanoOpticsBook, Brody2013, VortexBrewster, UB_NK_NKV_OL_2022, ADetalNormInc2023}, as shown in Fig. \ref{Fig_ExpSetup}. 
A Gaussian He-Ne laser beam ($\lambda = 632.8$ nm) from the source $L_S$ is passed through a collimating lens pair $L_P$. The resulting collimated beam is passed through a half wave plate (HWP) $H$ and a Glan-Thompson polarizer (GTP) $G_1$, via which 
an initial polarization $-\yhat$ is imparted (considering the source laser already has some polarization, the role of $H$ is to reorient it to maximize the $G_1$ output intensity). A beam splitter $B_S$ then partially reflects this beam towards a high NA lens $L$. The lens tightly converges the beam towards a mirror $M$. The mirror is placed on a translation stage so that it can be moved along the $z$ axis in the region of the focus. The resulting reflected beam passes again through the lens $L$, partially transmits again through $B_S$, and propagates towards a CCD camera $C_C$ for observation.

When placed exactly at the focus, an ideal mirror reflects the beam in such a way that the final output beam exiting the lens $L$ is purely $\yhat$ polarized (corresponding to the polarization $-\yhat$ input to $L$). This can be attributed to the Fresnel reflection coefficient values $r^\mathrm{p} = -r^\mathrm{s} = 1$ of an ideal mirror \cite{NanoOpticsBook}. So, it is important to use a mirror here instead of a dielectric or other special interface, as additional effects due to Fresnel coefficient variations and other special interface properties are removed.
This allows us to pay attention only to the wavefront curvature and diffraction effects, which aligns with our central objective and the choice of a simple initial field in the simulated model.

We denote the $z$ position of the mirror as $z_0$ [Fig. \ref{Fig_ExpSetup}], with $z_0 = 0$ being the exact focal position.
While the simulation results are demonstrated by varying the \roc $Z_C$, the experimental results are to be observed by varying $z_0$. So it is essential to understand, at least qualitatively, how $Z_C$ and $z_0$ are related. First, we see that the final output beam would not be collimated for $z_0 \neq 0$. This residual wavefront curvature is represented by the \roc $Z_C$. Next, it is easily verifiable by using a simple ray diagram that the output beam is converging for $z_0 < 0$ and diverging for $z_0 > 0$. This establishes the following correlations between $Z_C$ and $z_0$:
\benum
\item $z_0 = 0$ implies $Z_C \rightarrow \infty$.
\item As $z_0$ is taken from $0$ to a negative (or positive) value, $Z_C$ reduces to a finite value with $\alpha = 1$ (or $\alpha = -1$), implying convergence (or divergence).
\eenum
The establishment of a quantitative relation between $Z_C$ and $z_0$ is out of the scope of the present paper. But the above qualitative relations are sufficient to correlate the simulated and experimental results.

With the above understanding, we proceed and identify that
the reflected beam after passing through $L$ gives the $\bE_A$ field [Eq. (\ref{EA})], and the field which is detected at $C_C$ is the $\bE_D$ field [Eq. (\ref{ED})]. In order to implement Stokes parameter measurements to determine the polarization characteristics of $\bE_D$, or specifically of $\bE_{\perp D}$ [Eq. (\ref{EperpD})], we place a QWP $Q_S$ and a GTP $G_S$ before $C_C$ [Fig. \ref{Fig_ExpSetup}]. 
By using appropriate orientations of $Q_S$ and $G_S$, the intensity profiles $\cI_{xD}$, $\cI_{yD}$, $\cI_{\pm D}$ and $\cI_{d^\pm D}$ are observed [Sec. \ref{Subsec_ExpObservable}].

The Gaussian half-width of the collimated beam is set as $w_0 \approx 1.53$ mm.
We use a microscope objective lens here as the high NA lens $L$, manufactured by Edmund Optics, with the following specifications: focal length 1.60 mm, oil immersion NA 1.25, magnification 100X. It has an aperture radius $a = 3$ mm.
As the mirror $M$, we use a broadband dielectric mirror manufactured by ThorLabs (part no. BB1-E02), which offers $|r^\mathrm{p}| \approx |r^\mathrm{s}| > 0.99$ for a range of angles of incidence $[0\deg,45\deg]$ and a range of wavelengths $[400,750]$ nm. 
The translation stage used to move the mirror produces the smallest unit displacement of $10$ $\mu$m, but it is possible to approximately measure a $5$ $\mu$m displacement via eye estimation. The distance between the lens aperture and the sensor of the camera $C_C$ is set as $D = 760$ mm.

Though the dielectric mirror offers $|r^\mathrm{p}| \approx |r^\mathrm{s}| \approx 1$, a small difference is sufficient to add an 
unintended $\xhat$ polarization to the output even in the absence of wavefront curvature and diffraction. 
In addition, the $|r^\mathrm{p}| \approx |r^\mathrm{s}| > 0.99$ specification of the mirror is valid only upto an angle of incidence $45\deg$. Since we use a high NA lens, angles of incidence much higher than $45\deg$ are involved in our experiment, thus 
causing considerable deviations from an ideal scenario.
As a consequence, the contribution of the $E_{xD} \xhat$ component in the $\bE_{\perp D}$ field is much higher than what is theoretically expected. In particular, $E_{xD}$ is nonzero even for $z_0 = 0$, and is comparable to $E_{yD}$ for most part of the $z_0$ variation. However, the field properties discussed in Sec. \ref{Subsec_ZCvariation}--\ref{Subsec_AddExampleSim} are certainly identifiable in the experiment, as we demonstrate below.

\subsection{Variation of \textit{z}$_0$} \label{Subsec_z0varExp}

\begin{figure*}
\includegraphics[width = \linewidth]{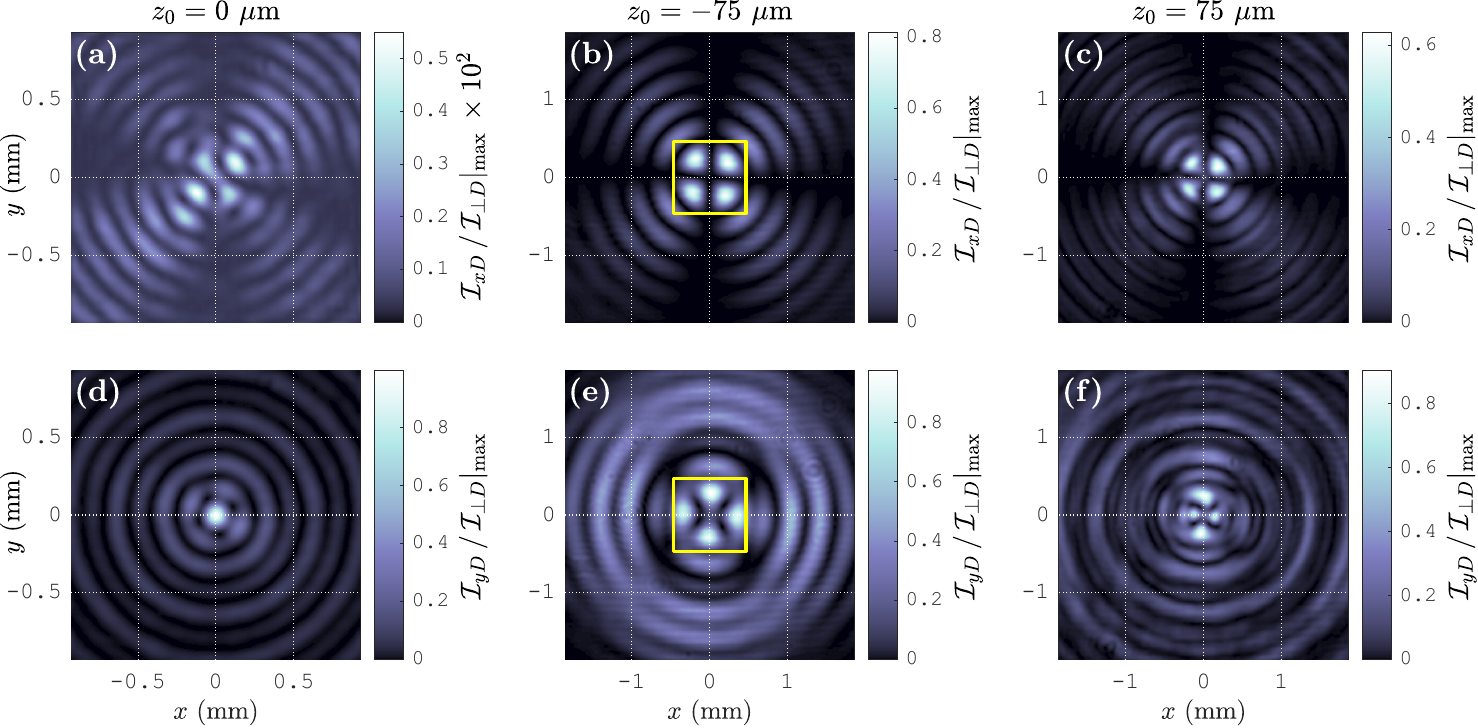}
\caption{
Experimental intensity profiles $\cI_{xD}$ and $\cI_{yD}$ for [(a),(d)] $z_0 = 0$ \mum, [(b),(e)] $z_0 = -75$ \mum, and [(c),(f)] $z_0 = 75$ \mum. 
The central region of interest, where the polarization characteristics are to be studied, is marked in (b) and (e).
}
{\color{grey}{\rule{\linewidth}{1pt}}}
\label{Fig_IxyD_Exp}
\end{figure*}

The experimentally observed $\cI_{xD}$ and $\cI_{yD}$ profiles for $z_0 = 0, \pm75$ $\mu$m are shown in Fig. \ref{Fig_IxyD_Exp}. Though the $\cI_{xD}$ intensity is experimentally nonzero for $z_0 = 0$ $\mu$m, its maximum value is two orders of magnitude less than that of $\cI_{yD}$. Correspondingly, the highly dominant $\cI_{yD}$ profile forms an approximate Airy pattern. But significant increase in the relative contribution of $\cI_{xD}$ in the total intensity is observed for $z_0 = \pm 75$ $\mu$m. These results show agreement with the simulated results of Sec. \ref{Subsec_ZCvariation}.

\begin{figure}
\includegraphics[width = \linewidth]{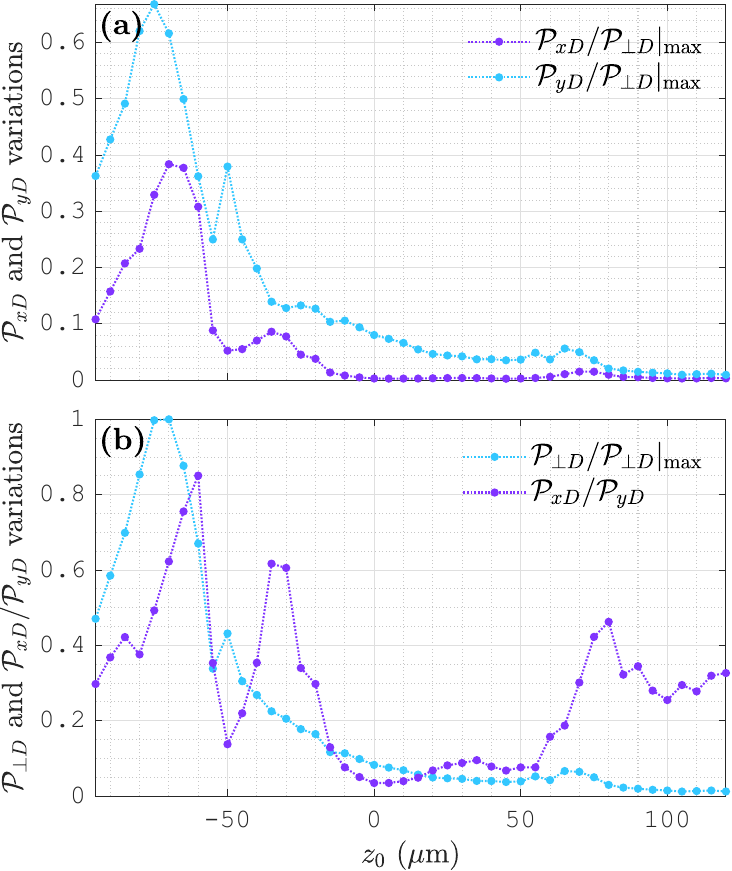}
\caption{
Experimentally observed variations of the powers $\cP_{x D}$, $\cP_{y D}$ and $\cP_{\perp D}$, and of the ratio $\cP_{x D} / \cP_{y D}$, \wrt the variation of $z_0$. The powers are normalized \wrt $\cP_{\perp D}|_\mathrm{max}$, i.e. the maximum value of $\cP_{\perp D}$.
}
{\color{grey}{\rule{\linewidth}{1pt}}}
\label{Fig_PwVar}
\end{figure}

To further understand these results, we observe the variation of the total power reaching the camera \wrt the variation of $z_0$. The powers are computed from the intensity data as 
\begin{\eq}
\cP = \int_{x_\mathrm{min}}^{x_\mathrm{max}} \int_{y_\mathrm{min}}^{y_\mathrm{max}} 
\cI \hpt dx \hpt dy.
\end{\eq}
The variations of the powers $\cP_{x D}$, $\cP_{y D}$ and $\cP_{\perp D}$, corresponding to the intensities $\cI_{x D}$, $\cI_{y D}$ and $\cI_{\perp D}$, are shown in Fig. \ref{Fig_PwVar}. The variation of the ratio $\cP_{x D} / \cP_{y D}$ is also shown in the figure. The plots show a few local maxima and minima which are presumably caused by the detailed mechanism of the microscope objective operation, and may not be explained 
quantitatively in this paper. 
But the global properties of these variations are clearly understood, as explained below. 

\benum

\item For all $z_0$, we get $\cP_{y D} > \cP_{x D}$, implying the dominance of the $E_{yD} \yhat$ component field over $E_{xD} \xhat$ in all cases.

\item An overall increase in the $z_0 < 0$ $\mu$m region and an overall decrease in the $z_0 > 0$ $\mu$m region are observed in $\cP_{y D}$ and $\cP_{\perp D}$. This signifies the increase in the detected power due to convergence, and the decrease in it due to divergence.

\item A global minimum of $\cP_{x D}$ is observed at $z_0 = 0$ $\mu$m, signifying the global minimum contribution of $E_{xD} \xhat$ in the total $\bE_{\perp D}$ field. While $\cP_{x D}$ increases on both sides of $z_0 = 0$ $\mu$m, the increase is much larger in the $z_0 < 0$ $\mu$m region due to convergence.

\item When $z_0$ is sufficiently far away from $0$ $\mu$m on both sides, all the powers decrease because of the eventual high divergence.

\eenum
These observations agree with the simulated results of Sec. \ref{Subsec_ZCvariation} reasonably well, and thus verify the correctness of the simulated as well as the experimental results.

It is important here to understand that 
it is extremely difficult to locate the $z_0 = 0$ $\mu$m position by manually seeing the mirror movement. 
This is essentially one of the core concerns we have begun with in Sec. \ref{Sec_Intro}.
So, for the present purpose, we have located the $z_0 = 0$ $\mu$m position in retrospect by identifying the global minima of $\cP_{x D}/\cP_{y D}$ in Fig. \ref{Fig_PwVar}. This method thus provides a significant and efficient way of locating the focus of a high NA lens in an actual experimental setup, and resolves the concern.

\subsection{Polarization Characteristics}

\begin{figure}
\includegraphics[width = \linewidth]{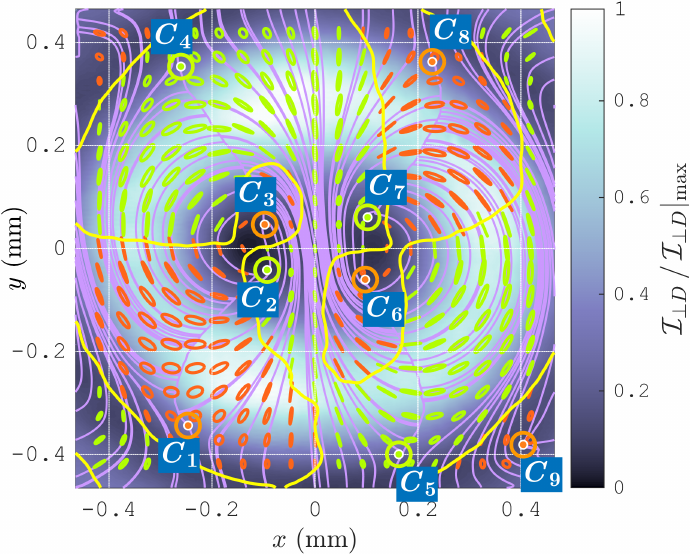}
\caption{
Experimentally obtained polarization profile of $\bE_{\perp D}$ for $z_0 = -75$ $\mu$m.
The $C$ point singularities $\{C_1,C_2,\cdots,C_9\}$ and the $L$ line singularities are identified. 
}
{\color{grey}{\rule{\linewidth}{1pt}}}
\label{Fig_EStreamExp}
\end{figure}

To observe the polarization properties of the $\bE_{\perp D}$ field, we perform Stokes parameter measurements for a few $z_0$ values, and construct the transverse field polarization profiles. For example, the $\bE_{\perp D}$ field profile for $z_0 = -75$ \mums is shown in Fig. \ref{Fig_EStreamExp}. This profile corresponds to the central regions of the $\cI_{xD}$ and $\cI_{yD}$ intensity profiles of Figs. \ref{Fig_IxyD_Exp}(b) and \ref{Fig_IxyD_Exp}(e). Nine $C$ point singularities are observed in the analyzed area, where $\{C_1, C_3, C_6, C_8, C_9\}$ are $\sighat^+$ polarized, and $\{C_2, C_4, C_5, C_7\}$ are $\sighat^-$ polarized. We also identify the boundaries between the REP and LEP regions as the $L$ line singularities. Though these lines are quite distorted because of the noise in the experimental data, their overall nature is clearly observed.

\begin{figure*}
\includegraphics[width = \linewidth]{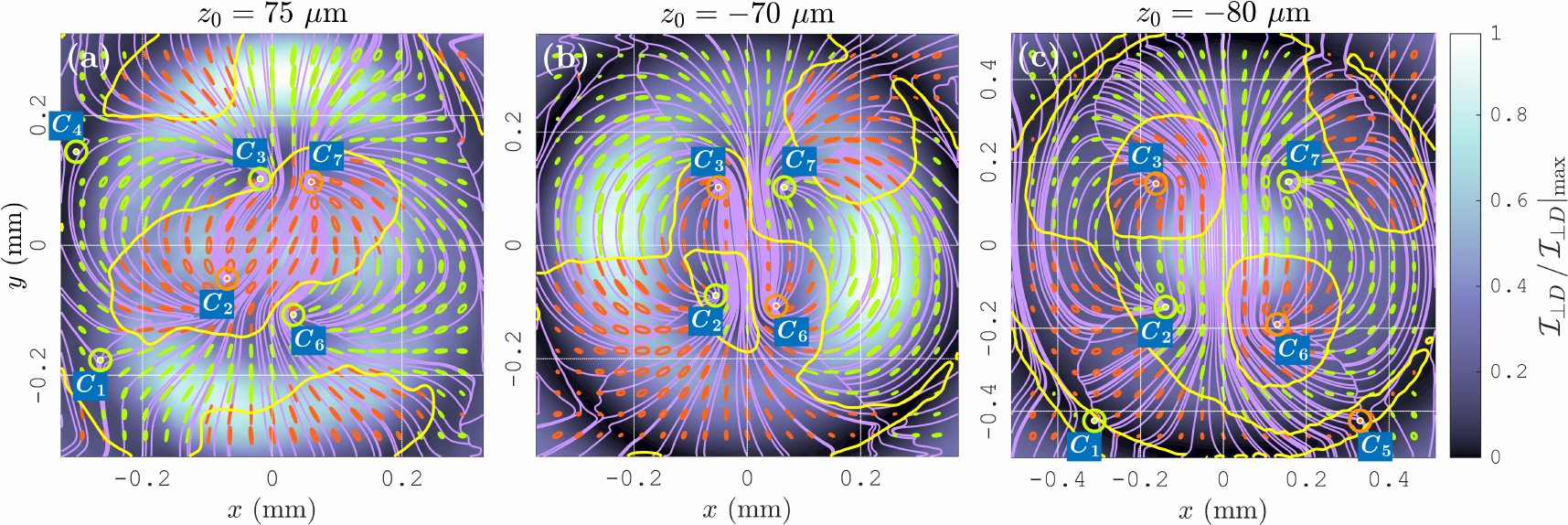}
\caption{
Experimentally obtained $\bE_{\perp D}$ profiles for some additional example cases. 
The $C$ points similar in nature to those in Fig. \ref{Fig_EStreamExp} are identified by the same names. Not all similar $C$ points appear (or are experimentally identified in the presence of noise) in all of the profiles, but the central points $\{C_2,C_3,C_6,C_7\}$ always appear. Any two adjacent central $C$ points have opposite spin polarizations, and these spins flip while transiting from the $z_0 < 0$ \mums region to the $z_0 > 0$ \mums region.
}
{\color{grey}{\rule{\linewidth}{1pt}}}
\label{Fig_EStreamExpExtra}
\end{figure*}

The polarization properties vary as $z_0$ is varied, as understood from the example $\bE_{\perp D}$ profiles shown in Fig. \ref{Fig_EStreamExpExtra}. The profile of Fig. \ref{Fig_EStreamExpExtra}(a) corresponds to the $\cI_{xD}$ and $\cI_{yD}$ profiles of Figs. \ref{Fig_IxyD_Exp}(c) and \ref{Fig_IxyD_Exp}(f), whereas, the profiles of Fig. \ref{Fig_EStreamExpExtra}(b) and \ref{Fig_EStreamExpExtra}(c) appear in the vicinity of the profile of Fig. \ref{Fig_EStreamExp}. Though the exact profile of $\bE_{\perp D}$ varies, the formation of the $C$ point and $L$ line singularities in all cases is clearly observed in these examples.

\begin{figure*}
\includegraphics[width = \linewidth]{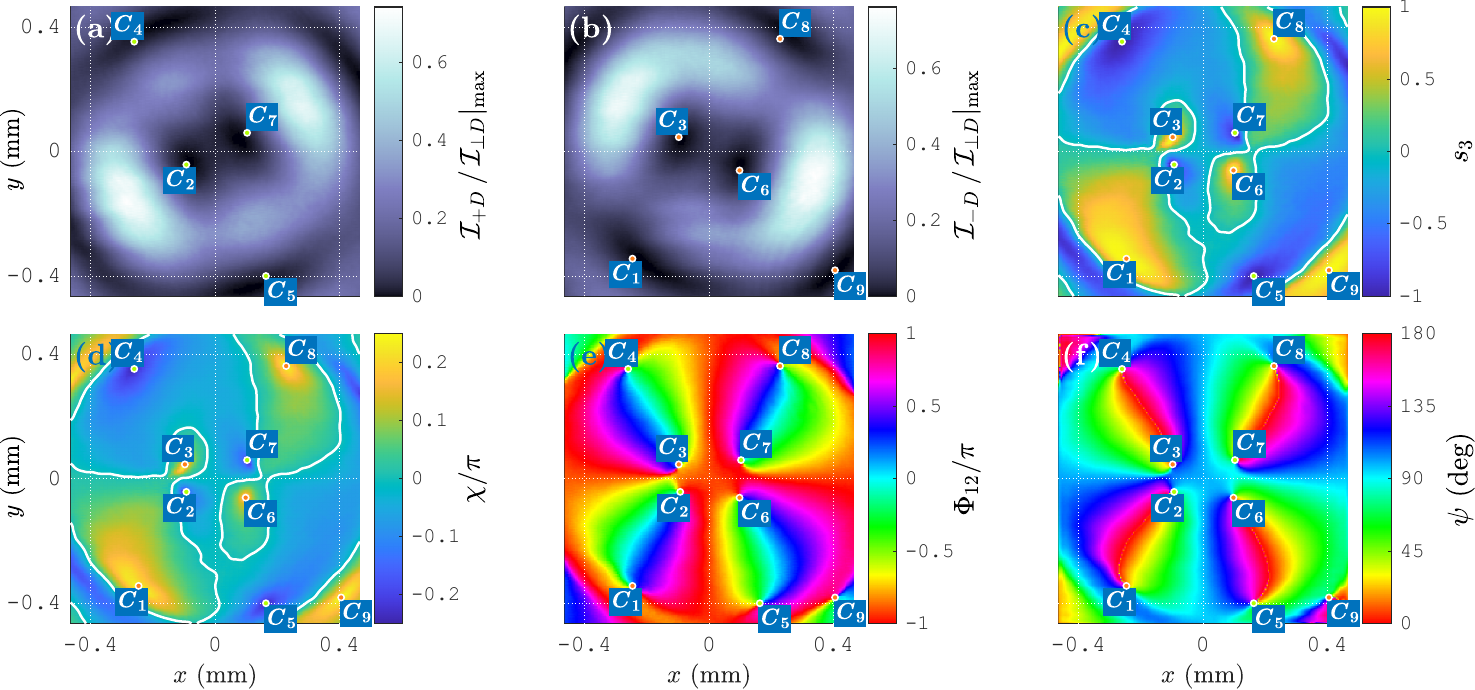}
\caption{
Experimentally obtained profiles of $\cI_{\pm D}$, $s_3$, $\chi$, $\Phi_{12}$ and $\psi$, corresponding to the $\bE_{\perp D}$ profile of Fig. \ref{Fig_EStreamExp}. 
The $C$ point singularities are marked. The $L$ line singularities ($s_3 = \chi = 0$) are shown in (c) and (d).
}
{\color{grey}{\rule{\linewidth}{1pt}}}
\label{Fig_StokesQuantitiesExp}
\end{figure*}

We can further understand the formation of these singularities by observing the intensities $\cI_{\pm D}$, normalized Stokes parameter $s_3$, ellipticity $\chi$, phase $\Phi_{12}$ and ellipse orientation $\psi$ [Sec. \ref{Subsec_ExpObservable}]. The profiles of these functions, considering the example of $z_0 = -75$ \mum, are shown in Fig. \ref{Fig_StokesQuantitiesExp}. It is easily observed that these profiles are related to the polarization singularities of Fig. \ref{Fig_EStreamExp} according to the relations established in Sec. \ref{Subsec_ExpObservable}.

In this context we see that, though the $\Phi_\pm$ phases [Eq. (\ref{EPhPM})] are not straightforwardly observed in the present experiment, their singularity characteristics can be interpreted by analyzing the experimental $\cI_{\pm D}$ and $\Phi_{12}$ profiles. A logical development of this understanding can be broken down into the following steps:

\benum

\item First, we interpret by observing the $\cI_\pm$ profiles that the $\Phi_+$ phase is singular at $\{C_2, C_4, C_5, C_7\}$, and the $\Phi_-$ phase is singular at $\{C_1, C_3, C_6, C_8, C_9\}$.

\item Because of the relation $\Phi_{12} = \Phi_- - \Phi_+$ [Eq. (\ref{Phi12})], there exists a singularity of $\Phi_{12}$ corresponding to each $\Phi_\pm$ singularity.
However, the correct sign must be considered while interpreting their topological charges. Because of the above relation, the topological charge of a $\Phi_-$ singularity and that of a corresponding $\Phi_{12}$ singularity have the same sign, but those of a $\Phi_+$ singularity and a corresponding $\Phi_{12}$ singularity have the opposite sign. 
This correlation can be clearly understood by comparing the simulated $\Phi_\pm$ and $\Phi_{12}$ profiles of Figs. \ref{Fig_abEPhiPM}(d), \ref{Fig_abEPhiPM}(h) and \ref{Fig_StokesQuantitiesSim}(e).
Thus, from the experimental $\Phi_{12}$ profile of Fig. \ref{Fig_StokesQuantitiesExp}(e), we interpret that the $\Phi_+$ singularities at $\{C_2, C_4, C_5, C_7\}$ have topological charges $\{-1, +1, +1, -1\}$, and the $\Phi_-$ singularities at $\{C_1, C_3, C_6, C_8, C_9\}$ have topological charges $\{-1, +1, +1, -1, +1\}$.

\item The formation of the lemon and star patterns of the streamlines in Fig. \ref{Fig_EStreamExp} can now be explained by using the above information, in the same way as in Sec. \ref{Subsec_CpointSim}.

\eenum


\section{Discussions and Future Directions} \label{Sec_Future}

The knowledge of the above simulated and experimental results helps us in finding ways to minimize the effects of the deviations due to wavefront curvature and aperture diffraction. Minimizing the distance $D$ between the lens aperture and the camera sensor [Fig. \ref{Fig_ExpSetup}] is the simplest way of reducing the effects of diffraction. But due to the placement of several optical components in between, this minimization faces restrictions. 
A better approach is to set the half-width $w_0$ sufficiently smaller than the aperture radius $a$. 
We see via Eqs. (\ref{circ_def}), (\ref{EA}) and (\ref{cE0Gauss}) that the amplitude of $\bE_A$ just inside the aperture boundary is 
$\cE_A(a) = \cE_{00} \hpt e^{-a^2/w_0^2} / \sqrt{\cos \theta_\mathrm{max}}$\,, 
whereas that everywhere outside the aperture boundary is zero. If the factor $e^{-a^2/w_0^2}$ is adjusted such that $\cE_A(a) \ll \cE_{00}$, then the contribution of $\circf(\rho/a)$ practically disappears, thus minimizing the effects of aperture diffraction. This can be achieved by making $w_0/a$ sufficiently smaller than unity in two possible ways: (1) by choosing a microscope objective with a large enough $a$, and/or (2) by adjusting the collimating lens pair $L_P$ to get a small enough $w_0$. 
The above description is in reference to our specific simulated model and experimental setup, but the general idea here is to achieve an appropriate underfilling of the lens aperture.
Equivalent conditions for other relevant optical systems can thus be implemented likewise.

The effect of wavefront curvature can be minimized with
a sufficiently accurate focal placement of the mirror (or any concerned interface or scatterer particle), which can be achieved 
in retrospect by observing the results of $z_0$ variation. One can first theoretically determine what effects are to be observed for $z_0 = 0$ and its variations, and subsequently identify these effects in the experimental observations to identify the $z_0 = 0$ position. In Sec. \ref{Subsec_z0varExp} we have elaborated this process for our present optical system, and one can also introduce customized prescriptions on possible observations for other optical systems likewise.

We have begun the present work in an attempt to understand the deviations which affect all optical systems involving the collimation of a reflected, transmitted or scattered wave subsequent to tight focusing. 
But at this point, the mathematical rigor involved in the system modelling, followed by the revelation of rich beam field characteristics and optical singularities, appears to elevate the significance of the present work far beyond a casual deviation analysis. After observing the results of Sec. \ref{Sec_Sim} and \ref{Sec_Exp}, we realize that these deviations are fundamentally significant optical phenomena by themselves.
The complex field profiles obtained here have remarkable similarities with the fields obtained at a tight focus in standard analyses \cite{RichardsWolf1959, NanoOpticsBook}. However, in our case these fields are not obtained at a tight focus, but instead, they are observed at a distant detector. The only common aspect between our obtained fields and the standard tight focal fields is the diffraction of a curved wavefront beam due to overfilling of the lens aperture. This aspect thus reveals that a complex field of this kind is not necessarily a tight focal field, but a general occurrence for any system where a highly diverging or converging beam is limited by a circular aperture.
With this remarkable realization, our paper already serves the purpose of a core level in-depth exploration of the concerned optical phenomena and complex fields.
Now in addition, we summarize here some future research areas where our present work can lead to regarding further explorations of this special class of optical processes.

\benum

\item One can build the experimental setup by using dielectric or other special surfaces in place of the mirror $M$. 
Correspondingly, in the simulated model one must replace the initial field $\cE_0 \yhat$ with an appropriate field determined by taking into account the surface properties. In this way, one can understand how the surface properties affect the observed fields, and in retrospect one can also device ways to identify the surface properties by observing the modified fields.

\item The evolution of the diffraction fringes can be studied by varying $w_0$ for a fixed $a$, i.e. by making transitions between the underfilling and overfilling conditions.

\item
If the distance $D$ is varied for a fixed $z_0$, the field profile evolves as the singularities move around in the beam cross section. This effect is equivalent to the vortex transformations and dynamics due to beam propagation \cite{MTerrizaChapterOAM2013}. 
In our experimental results we have kept $D$ fixed and have varied $z_0$, which also has resulted into similar transformations of the singularities. To interpret both of these phenomena in a general mathematical form, we can re-express $\bE_{\perp D}$ by using Eqs. (\ref{EperpD_sigmaPM}) and (\ref{EPM_abPM_full}) as
\begin{\eq}
\bE_{\perp D} = |E_{+ D}| \hpt e^{i \Phi_+} \sighat^+ + |E_{- D}| \hpt e^{i \Phi_-} \sighat^- .
\end{\eq}
This is a nonseparable state representation, which describes a coupling between the spin states $\sighat^\pm$ and the spatial states $|E_{\pm D}| \hpt e^{i \Phi_\pm}$. Using the states $|E_{\pm D}| \hpt e^{i \Phi_\pm}$ one can determine the orbital angular momenta (OAM) associated to the spin states $\sighat^\pm$ \cite{AllenOAM1992, Berry1998, Soskin1997, Barnett2002, DovePrisms2002, TriAperture2010, SchulzeModalDecomp2013, LaveryC13_OAM2013, Zhang2015, Gbur, ADNKVSPIE2023}, thus characterizing the SOC in the beam field. This coupling evolves due to the variation in the source field $\bE_A$ (achieved via $z_0$ variation) as well as due to the beam propagation (observed via $D$ variation). 
A detailed exploration of this SOC evolution can be carried out in the future.
One can anticipate to observe an exchange of OAM between the $\sighat^\pm$ spin polarized fields due to this SOC evolution, while the total angular momentum of the beam field would remain conserved.

\item 
Figures \ref{Fig_EstreamSim}, \ref{Fig_EStreamSimExtra}, \ref{Fig_EStreamExp} and \ref{Fig_EStreamExpExtra} show the $\bE_{\perp D}$ field profile in very small regions around the beam center. Clearly, the entire beam field contains a large number of optical singularities. The complexity of the $\bE_{\perp D}$ field may remind one of the singularities of speckle fields \cite{GoodmanSpeckle}. However, the $\bE_{\perp D}$ field is fundamentally different from a speckle field because the latter possesses an inherent randomness, 
whereas the former originates from a well defined source field $\bE_{A}$ [Eq. (\ref{EA})] and is well defined via Eqs. (\ref{ED}) and (\ref{EperpD}).
Since $\bE_{\perp D}$ is not random, recognizable common characteristics of the singularities and the streamline patterns are observed in all the profiles of Figs. \ref{Fig_EstreamSim}, \ref{Fig_EStreamSimExtra}, \ref{Fig_EStreamExp} and \ref{Fig_EStreamExpExtra}, which is usually not the case for speckle fields.
Such similarities and dissimilarities of the $\bE_{\perp D}$ field with speckle fields can be explored in the future.

\item As the observed singularities can be manipulated by varying $z_0$, their applications in optical tweezing and particle rotation is a possibility that can be explored in the future.

\item Finally, since complex fields of the similar nature appear also in other systems involving diffraction-limited highly diverging and converging beams, similar field analyses can be performed in all such systems as well. This class of systems and processes may include the standard tight focusing, the `deviation from collimation' kind considered in this paper, and any other conceivable system where a highly diverging or converging beam is passed through a circular aperture. The future directions mentioned in the previous points can also be studied considering appropriate designs of all such systems.

\eenum




\section{Conclusion} \label{Sec_Conc}

To summarize, we have described two kinds of deviations which affect the study of optical fields when a tightly focused beam is reflected, transmitted or scattered, and subsequently collimated: one is due to wavefront curvature, and the other is due to aperture diffraction. We have simulated an appropriate model optical system to understand these effects. 
In the simulation we have varied the wavefront curvature, 
and have used FFT algorithm to compute the passing of the resulting field through a circular aperture. We have thus demonstrated the complicated intensity profiles and optical singularity characteristics of the far field.

Subsequently, we have built an experimental setup in the form of a normal incidence and reflection system using a microscope objective lens. In the setup we have achieved the wavefront curvature variation by displacing the reflecting mirror. The experimentally observed intensity and optical singularity characteristics agree well with the simulated results, and thus verify the correctness of our analysis.

Finally, based on our analysis we have described how to minimize the deviations in differently purposed experiments, which would be relevant to novel applications such as material characterization, dark field microscopy, nanoprobing etc. 
But the most important outcome of the present paper is that the identified deviations are understood to be significant electromagnetic optical phenomena by themselves. 
This aspect is clearly revealed by the rigorous system modelling and the subsequent detailed simulated and experimental results on the beam field profiles.
While the present analysis and results already serve as core level explorations of these phenomena, 
we have also listed a few directions in which one can carry out investigations to further explore some of their other significant characteristics and future applications.

\begin{acknowledgments}
We thank Upasana Baishya (School of Physics, University of Hyderabad) for valuable relevant discussions.
N.K. thanks University Grants Commission (India) for Institutions of Eminence (IoE) Incentive.
A.D. thanks Council of Scientific and Industrial Research (India) for Senior Research Fellowship (SRF). N.K.V. thanks Science and Engineering Research Board (Department of Science and Technology, India) for financial support.
\end{acknowledgments}


\bibliography{NK_AD_NKV_NormIncDiffraction_Refs}

\end{document}